\documentclass[
preprint, 
secnumarabic,amssymb, nobibnotes, aps, prd
,superscriptaddress,
nofootinbib
]{revtex4-1}

\usepackage[ddmmyyyy]{datetime} 
\usepackage{amsmath,color,xcolor} 
\usepackage{graphicx} 
\usepackage{float}
\usepackage{soul,ulem} 

\newcommand{\md}{\mathrm{d}}
\def\be{\begin{equation}}
\def\ee{\end{equation}}

\definecolor{brightpink}{rgb}{1.0, 0.0, 0.5}

\begin{document}

\title{Hamiltonian analysis of mimetic scalar gravity revisited}%

 \author{Alexander Ganz}
 \email{alexander.ganz@pd.infn.it}
 \author{Purnendu Karmakar}
 \email{purnendu.karmakar@pd.infn.it}
 \affiliation{ Dipartimento di Fisica e Astronomia ``G. Galilei", \\
	Universit\`a degli Studi di Padova, via Marzolo 8, I-35131 Padova, Italy}
 \affiliation{INFN, Sezione di Padova, via Marzolo 8, I-35131 Padova, Italy}
 \author{Sabino Matarrese}
 \email{sabino.matarrese@pd.infn.it}
 \affiliation{ Dipartimento di Fisica e Astronomia ``G. Galilei", \\
	Universit\`a degli Studi di Padova, via Marzolo 8, I-35131 Padova, Italy}
 \affiliation{INFN, Sezione di Padova, via Marzolo 8, I-35131 Padova, Italy}
 \affiliation{INAF - Osservatorio Astronomico di Padova, Vicolo dell'Osservatorio 5, I-35122 Padova, Italy}
\affiliation{Gran Sasso Science Institute, Viale F. Crispi 7, I-67100 L'Aquila, Italy}
 \author{Dmitri Sorokin}
 \email{dmitri.sorokin@pd.infn.it}
\affiliation{INFN, Sezione di Padova, via Marzolo 8, I-35131 Padova, Italy} \affiliation{ Dipartimento di Fisica e Astronomia ``G. Galilei", \\
	Universit\`a degli Studi di Padova, via Marzolo 8, I-35131 Padova, Italy}


\begin{abstract}
We perform the Hamiltonian analysis of several mimetic gravity models and compare our results with those obtained previously by different authors.  We verify that for healthy mimetic scalar-tensor theories the condition for the corresponding part of the Hamiltonian to be bounded from below is the positive value of the mimetic field energy density $\lambda$. We show that for mimetic dark matter possessing a shift symmetry 
the mimetic energy density remains positive in time, provided appropriate boundary conditions are imposed on its initial value, while in models without shift symmetry the positive energy density can be maintained by simply replacing $\lambda\to e^{\lambda}$. The same result also applies to mimetic $f(R)$ gravity, which is healthy if the usual stability conditions of the standard $f(R)$ gravity are assumed and $\lambda>0$.
In contrast, if we add mimetic matter to an unhealthy seed action, the resulting mimetic gravity theory remains, in general, unstable. As an example, we consider a scalar-tensor theory with the higher-derivative term $(\Box \varphi)^2$, which contains an Ostrogradski ghost.
We also revisit results regarding stability issues of linear perturbations around the FLRW background of the mimetic dark matter in the presence of ordinary scalar matter. We find that the presence of conventional matter does not revive dynamical ghost modes (at least in the UV limit). The modes, whose Hamiltonian is not positive definite, are non-propagating (have zero sound speed) and are associated with the mimetic matter itself. They are already present in the case in which the ordinary scalar fluid is absent, causing a growth of dust overdensity. 
\end{abstract}


\date{\today}%

\maketitle

\section{Introduction}
As was realized some time ago \cite{Lim:2010yk,Gao:2010gj,Capozziello:2010uv,Chamseddine:2013kea}, a cold component of dark matter (and dark energy) may be mimicked by adding to General Relativity (GR) a ``perfect fluid of dust". Such a mimetic dark matter can be introduced \cite{Chamseddine:2013kea} by performing a non-invertible conformal transformation of the GR metric, with a re-scaling parameter being a kinetic term for a (mimetic) scalar field
\begin{equation}\label{ni1}
g_{\mu\nu} = - \left(\tilde{g}^{\alpha \beta}\partial_\alpha \varphi \,. \partial_\beta\varphi\right) \tilde{g}_{\mu\nu}\,.
\end{equation}
The transformation is non-invertible in the sense that the metric $\tilde g_{\mu\nu}$ cannot be fully expressed in terms of $g_{\mu\nu}$ and $\varphi$ due to the invariance of the right-hand side of \eqref{ni1} under the conformal re-scaling of the metric $\tilde g_{\mu\nu}$.
In this way one obtains a conformally invariant theory. Later on, such a non-invertible transformation (or its generalizations) has been applied to wide classes of models including general scalar-tensor theories, $f(R)$-gravity, vector-tensor theories etc., resulting in mimetic versions of these models \cite{Barvinsky:2013mea,Chamseddine:2014vna,Chaichian:2014qba,Nojiri:2014zqa,Arroja:2015wpa,Arroja:2015yvd,Vikman:2017gxs,Kluson:2017iem,Takahashi:2017pje,Langlois:2018jdg,Jirousek:2018ago,Nojiri:2018ouv}, which is usually referred as `mimetic gravity'.

A natural question which arises is whether or not the addition of mimetic matter may cause or cure instabilities in these models. This problem has been studied in different models by various authors. For instance, for the simplest mimetic dark matter model, it was shown \cite{Barvinsky:2013mea,Chaichian:2014qba} that for the system to be free of ghost instabilities the necessary condition is that the energy density of the mimetic dust should remain positive definite under time evolution, but that this may not always be the case. Refs. \cite{Langlois:2018jdg,Takahashi:2017pje,Zheng:2017qfs,Firouzjahi:2017txv,Ijjas:2016pad,Hirano:2017zox,Gorji:2017cai} studied the behavior of a wide class of mimetic gravity models under linear perturbations around certain cosmological backgrounds (such as Friedmann-Lema\^itre-Robertson-Walker - FLRW - ones) and argued that they might be plagued with ghost-like and/or gradient instabilities. In particular \cite{Takahashi:2017pje,Langlois:2018jdg} observed a ghost-like instability of linear perturbations around the FLRW background in the presence of conventional scalar matter in a unitary gauge in which the mimetic scalar is identified with the time flow. In this respect, it is important to understand if the source of the instability is the presence of the additional matter itself or whether it is intrinsic to the mimetic field itself. For instance, the Jeans instability is well-known to appear around certain GR backgrounds in the presence of conventional matter \cite{Jeans1902,1946ZhETF..16..587L}. 

The main aim of this paper is to revisit these issues. 
To check, in full generality, whether the mimetic field can bring additional instabilities into a given gravity model one should first perform a complete background independent Hamiltonian analysis of a full non-linear system and check under which conditions different parts of the Hamiltonian are bounded from below, which ensures the existence of stable solutions. The Hamiltonian describing a wide class of mimetic gravity models was derived in \cite{Takahashi:2017pje}, together with first and second class constraints which allowed to count the number of physical degrees of freedom. However, the stability analysis of this Hamiltonian has not been carried out therein. The full Hamiltonian analysis for a simple mimetic dark matter model was carried out in \cite{Chaichian:2014qba,Malaeb:2014vua} and extended in \cite{Kluson:2017iem} to mimetic gravity actions involving an arbitrary function $F(\Box \varphi)$ of the mimetic scalar field and further generalized in \cite{Zheng:2018cuc}. As we already mentioned, the main conclusion was that the stability of the mimetic system requires that the energy density of the mimetic dust is non-negative. In this paper we will extend these results to a wider class of mimetic gravity models including conventional matter and also study the stability of their linear fluctuations around the FLRW background, revisiting results of \cite{Takahashi:2017pje,Langlois:2018jdg}.

In Section \ref{sec:Hamiltonian_analysis_DHOST} we analyze the stability conditions for mimetic scalar-tensor theories by performing the full Hamiltonian analysis and verifying that in all the cases a necessary condition for the corresponding part of the Hamiltonian to be bounded from below is the positive value of the mimetic field energy density $\lambda$. We will show that for the mimetic dark matter model possessing a shift symmetry $(\varphi\to\varphi+c)$ the mimetic energy density remains positive in time, provided appropriate boundary conditions are imposed on its initial value. For more general cases (without shift symmetry) the positivity condition on mimetic energy density may be imposed {\it a priori} by requiring, e.g. that $\lambda=e^{\hat\lambda}$ with $\hat\lambda$ being an arbitrary scalar field.

In Section \ref{FLRWB} we will revisit results of \cite{Takahashi:2017pje,Langlois:2018jdg}, where it was observed (using the unitary gauge $\varphi=t$) that, in the presence of matter, linear perturbations of mimetic gravity models around the FLRW background have an Ostrogradski ghost.
To better understand the nature of this instability we will derive the corresponding second-order action and Hamiltonian without gauge fixing local reparametrization invariance and give the result in terms of gauge-invariant variables. For comparison, we will also present the results of the analysis of linear perturbations for the pure mimetic matter, in the absence of conventional matter, and vice versa. We will see, taking an ultra-violet limit, that the presence of conventional matter does not revive dynamical ghost modes. The modes, whose Hamiltonian is not positive definite, are non-propagating (with zero sound speed) and are associated with the mimetic matter itself, as in the case in which the conventional scalar fluid is absent. Though, as we will show, at the linearized level one cannot unambiguously identify the nature of these modes, i.e., whether they are ghost-like or tachyon-like, we will see that they cause the usual Jeans instability of dust. On top of these, the conventional matter brings about two propagating modes with positive definite Hamiltonian. 

In Section \ref{sec:Higher_Derivative_terms} we consider a mimetic theory based on an unhealthy primary seed action with a higher-derivative term $\left(\Box \varphi\right)^2$ and show that the mimetic constraint cannot, in general, cure instability problems of the primary action. In Section \ref{sec:mimetic_f(R)} we will also shortly discuss the stability conditions, due to the mimetic constraint, for a bounded Hamiltonian of the mimetic $f(R)$ gravity. Finally, in the Conclusions, we present a short summary of our analysis. Some calculations are given in the Appendix. 

In our paper we are using the $(-,+,+,+)$ signature for the metric. Greek indices run from 0 to 3 and Latin indices from 1 to 3. Further, we are working in units where the speed of light and the reduced Planck mass is one.


\section{A brief survey of the structure of mimetic gravity models, also in view of frames imposed by the GW170817 event}
\label{sec:Equivalence_approaches}
As largely discussed in the literature \cite{Deruelle:2009pu,Chamseddine:2013kea,Barvinsky:2013mea,Chamseddine:2014vna,Chaichian:2014qba,Arroja:2015wpa,Takahashi:2017pje,Langlois:2018jdg}, mimetic gravity can be obtained upon performing a disformal transformation \cite{Bekenstein:1992pj} of the metric in an original theory and then requiring that this transformation is non-invertible. Such a non-invertible disformal transformation can be written as a non-invertible conformal transformation followed by an invertible disformal transformation \cite{Takahashi:2017pje,Langlois:2018jdg}. The invertible disformal transformation does not change the physical content of the theory, but the non-invertible conformal transformation does change the theory, giving rise to a mimetic gravity. Therefore, to arrive at mimetic gravity from a given gravity model, one can always choose the disformal transformation to be a Weyl transformation of the metric
\begin{equation}\label{Weyl}
g_{\mu\nu}(x)=\tilde X(x)\tilde g_{\mu\nu}(x)\,,
\end{equation}
where $\tilde X(x)$ is the rescaling parameter. Upon this transformation, a generic Lagrangian
\begin{equation}\label{Lg}
\mathcal L=\sqrt{-g}f(\eta,\varphi)R+\mathcal L_{matter}(g_{\mu\nu},\eta,\varphi)\,,
\end{equation}
including gravity, a scalar field $\varphi$ and other matter fields $\eta$, takes the following form
\begin{equation}\label{XLg}
\tilde{\mathcal L}=\sqrt{-\tilde g}f(\eta,\varphi,\tilde X)\Big(\tilde X\tilde R+\frac 32\tilde X^{-1}\tilde g^{\mu\nu}\partial_\mu\tilde X\partial_\nu\tilde X-3\Box \tilde X\Big)+\mathcal L_{matter}( \tilde g_{\mu\nu},\eta,\varphi,\tilde X)\,,
\end{equation}
where we assumed that the fields $\varphi$ and $\eta$ do not transform under the Weyl transformation.
The generic $\tilde X$ is a non-dynamical St\"uckelberg-like field which can be gauge fixed to a constant by the inverse Weyl transformation upon which we get back the initial Lagrangian (\ref{Lg}). In other words, as is well known, the Lagrangian (\ref{XLg}) is invariant under the local conformal (Weyl) transformations
\begin{equation}\label{Weyl1}
\tilde g_{\mu\nu} \to \Omega(x)\tilde g_{\mu\nu}\,, \qquad \tilde X \to \Omega^{-1}(x)\tilde X\,.
\end{equation}

Let us now assume that $\tilde X(x)$ is not an independent field but is proportional to the kinetic term of $\varphi$ which may be (or not) part of the Lagrangian $\mathcal L_{matter}(\tilde g_{\mu\nu}, \varphi,\eta,\tilde X)$
\begin{equation}\label{Xdfdf}
\tilde X=-\tilde g^{\mu\nu} \partial_\mu\varphi\partial_\nu\varphi\,.
\end{equation}
If we substitute the expression (\ref{Xdfdf}) into (\ref{XLg}) we see that the field $\varphi$ contributes to the Lagrangian with higher-derivative terms. The Lagrangian is still invariant under the Weyl rescaling (\ref{Weyl1}) of the metric $\tilde g_{\mu\nu}$, but now if we gauge fix the Weyl symmetry by putting $\tilde X=1$ we get the mimetic constraint on the kinetic term of $\varphi$
\begin{equation}\label{mimc}
\tilde g^{\mu\nu} \partial_\mu\varphi\partial_\nu\varphi=-1\,.
\end{equation}
This implies that with the specific choice of $\tilde X$ as in (\ref{Xdfdf}) the theory has a constrained mimetic scalar degree of freedom $\varphi$. This gauge fixing can also be understood as a field redefinition \cite{Hammer:2015pcx} (associated with a Weyl transformation) which eliminates the St\"uckelberg field as follows. Let us, following \cite{Chaichian:2014qba}, introduce the relation \eqref{Xdfdf} into the Lagrangian \eqref{XLg} as a Lagrange multiplier term
\begin{equation}
{\mathcal L}_\lambda=\sqrt{-\tilde g}\tilde\lambda(\tilde X+\tilde g^{\mu\nu} \partial_\mu\varphi\partial_\nu\varphi)\,.
\end{equation}
Now let us make the following field redefinition ${\tilde g}_{\mu\nu}={\tilde X}^{-1}g_{\mu\nu},\; \tilde\lambda={\tilde X}\lambda$ upon which the Lagrangian $\tilde{\mathcal L}+\mathcal L_\lambda$ reduces to
\begin{equation}\label{l+l}
\mathcal L=\sqrt{-g}f(\eta,\varphi)R+\mathcal L_{matter}(g_{\mu\nu},\eta,\varphi)-\sqrt{-g}\lambda(1+ g^{\mu\nu} \partial_\mu\varphi\partial_\nu\varphi)\,,
\end{equation}
which produces the mimetic constraint \eqref{mimc}. Therefore, when analyzing the theory, one can keep Weyl invariance to a certain point and gauge fix it by imposing \eqref{mimc} at a later stage, as e.g. in \cite{Chaichian:2014qba,Takahashi:2017pje}. 

Alternatively, one can from the beginning add the mimetic constraint \eqref{mimc} to the initial Lagrangian \eqref{Lg} as a Lagrange multiplier term \cite{Golovnev:2013jxa} and get \eqref{l+l}.  Note that in the mimetic dark matter model \cite{Chamseddine:2014vna} the field $\lambda$ has actually a clear physical meaning of being associated with mimetic matter energy density. If the initial action in \eqref{Lg} contains the kinetic term of the scalar field $X\equiv- g^{\mu\nu}\partial_\mu\varphi\partial_\nu\varphi$ and/or the functions $f$, $L_{matter}$ etc. depend on $X$, this dependence can always be removed by expanding these functions \footnote{$f(X)=f( X-1+1)=f|_{ X-1}+(X-1)f'+\frac 12( X-1)^2f''+\cdots$} in powers of $ X-1$ and adsorbing the $( X-1)$-dependent terms by the Lagrange multiplier $\lambda$. Therefore, as discussed in \cite{Langlois:2018jdg}, every higher derivative term of the scalar field $\varphi$ in the initial Lagrangian, which contains a covariant derivative of $X$ will effectively vanish upon the non-invertible conformal transformation leading to mimetic gravity. Hence, the essential higher-derivative terms in the initial action (depending on the second derivative of $\varphi$) which survive upon the disformal Weyl transformation are of the following 
schematic form $\chi_n = g^{\mu\nu} [ \varphi]_{\mu\nu}^n$ \footnote{Here and in what follows we use the notation of \cite{Langlois:2018jdg}, where $n$ is the polynomial order of second order derivatives, such that $\varphi_\mu \equiv \partial_\mu \varphi$, $\varphi_{\mu\nu}= \nabla_\mu \partial_\nu\varphi$,
$\chi_1= g^{\mu\nu} \varphi_{\mu\nu}$, $\chi_2= \varphi_{\mu\nu} \varphi^{\mu\nu}$, $\chi_3 = \varphi_{\mu\nu} \varphi^{\mu\rho} \varphi_{\rho}^{\nu}$ etc.}.
Consequently, the most general scalar-tensor mimetic theory action (depending on up-to-second-order derivatives of $\varphi$) can be taken in the form
\begin{align}\label{Sn}
\mathcal L=\sqrt{-g}f(\eta,\varphi)R+ \mathcal{L}_\varphi(\varphi,\eta,\chi_1,...,\chi_n) + \mathcal L_{matter}(g_{\mu\nu},\eta,\varphi)- \sqrt{-g}\lambda\left(g^{\mu\nu}\partial_\mu\varphi\partial_\nu\varphi + 1\right)\,.
\end{align}
\if{}
Upon the disformal Weyl transformation it becomes
\begin{align}\label{Sn}
 \tilde{ \mathcal L}=\sqrt{- \tilde g}f(\eta,\varphi) \tilde R+ \tilde{\mathcal{L}}_\varphi(\varphi,\eta,\tilde \chi_1,...,\tilde \chi_n) + \mathcal L_{matter}(\tilde g_{\mu\nu},\eta,\varphi) \,,
\end{align}
where we imposed the mimetic constraint by fixing the Weyl invariance and introducing the Lagrange multiplier term.
At this point, the Lagrangian depends only on the metric $\tilde g_{\mu\nu}$, and in what follows we will simply remove ``tilde" 
from all the expressions.
\fi

The recent observation, from the binary neutron star merging event GW170817 \cite{TheLIGOScientific:2017qsa}, that the speed of gravitational waves is equal to the speed of light, to a very high accuracy, restricts the class of viable healthy seed models. In the absence of conventional matter, these are described by the action \cite{Ganz:2018vzg}
\begin{equation}
\label{eq:Hamiltonian_analysis_action}
S = \int\md^4x\, \sqrt{-g}\,\left( f(\varphi)R - V(\varphi) \right) - \int\md^4x\, \sqrt{-g}\, \lambda \left(g^{\mu\nu}\partial_\mu\varphi\partial_\nu\varphi + 1\right)\,.
\end{equation}
This model is usually referred to as `generalized mimetic scalar tensor theory', while the original `mimetic dark matter' model is a special case of the above action in which $f=1/2$ and $V=0$.

To the action (\ref{eq:Hamiltonian_analysis_action}) we can add terms of the form $F(\Box \varphi)$, which do not change the propagation speed of the gravitational waves. While terms proportional to $\Box \varphi$ can be reabsorbed by a redefinition of $V(\varphi)$ (see the discussion in \cite{Ganz:2018vzg}), higher-derivative terms, such as $(\Box\varphi)^2$, or in general $F(\Box \varphi)$, are commonly discussed in the literature on mimetic gravity \cite{Chamseddine:2014vna,Mirzagholi:2014ifa,Ijjas:2016pad,Zheng:2017qfs,Firouzjahi:2017txv,Kluson:2017iem}. Terms of this kind are interesting, since they provide a non-vanishing sound speed \cite{Chamseddine:2014vna}, while in mimetic Horndeski models the sound speed vanishes \cite{Arroja:2015yvd}. Further, they provide an interesting connection to the infrared limit of the projectable Ho\v rava-Lifshitz gravity \cite{Mukohyama:2009mz,Blas:2009yd,Ramazanov:2016xhp}. However, in general, seed actions containing terms of this kind contain an Ostrogradski ghost, and, as we will see in section \ref{sec:Higher_Derivative_terms}, with the example of action 
\begin{equation}\label{Sbox}
    S =\frac{1}{2} \int \md^4x\,\sqrt{-g}\, R + \int \md^4x\,\sqrt{-g}\, \left( G(\varphi) (\Box \varphi)^2 -\lambda \left(g^{\mu\nu}\partial_\mu\varphi\partial_\nu\varphi + 1\right)\right) \;,
\end{equation}
the mimetic constraint, though it reduces by one the number of independent degrees of freedom, it does not solve instability issues.

\section{Hamiltonian analysis of the mimetic scalar-tensor theory}
\label{sec:Hamiltonian_analysis_DHOST}
In this section, we discuss the stability conditions for the mimetic scalar-tensor theory described above in \eqref{eq:Hamiltonian_analysis_action}, by performing the full Hamiltonian analysis and checking whether the Hamiltonian is bounded from below. 

In order to simplify calculations, we will perform the Hamiltonian analysis of the theory with the mimetic constraint introduced via the Lagrange multiplier term. As such, we will not deal with a first-class constraint associated with the Weyl symmetry as in \cite{Takahashi:2017pje,Chaichian:2014qba}, but instead with two second-class constraints associated with the presence of the Lagrange multiplier, and the mimetic constraint as the gauge-fixing condition.

The main goal is to analyze the stability conditions associated with the presence of the mimetic matter and compare them with those in the corresponding ``non-mimetic" models. 

\subsection{ADM decomposition}
To set our notation and conventions, we start with a review of the well-known techniques for carrying out the Hamiltonian analysis of the theories involving gravity.

For the foliation of spacetime we are using the ADM decomposition with metric
\begin{equation}
\md s^2 = - N^2 \md t^2 + h_{ij} \left( \md x^i + N^i \md t\right) \left( \md x^j + N^j \md t\right)\,,
\end{equation}
where $N$ is the lapse, $N^i$ is the shift vector, and $h_{ij}$ is the three dimensional metric of the hypersurface of constant time which is used to raise and lower spatial indices. The Ricci scalar of the four dimensional metric $g_{\mu\nu}$ can be decomposed as
\begin{equation}
R = \bar{R} + K_{ij}K^{ij} - \left(K_i^i\right)^2 + 2\, \nabla_\mu \left( n^\mu \nabla_\nu n^\nu -n^\nu \nabla_\nu n^\mu\right)\,,
\end{equation}
where $\bar{R}$ is the three dimensional Ricci scalar, and $K_{ij}$ is the extrinsic curvature
\begin{equation}\label{Kij}
K_{ij}=\frac{1}{2N}\left(\dot{h}_{ij}- 2 D_{(i}N_{j)}\right)\,.
\end{equation}
where $2 D_{(i}N_{j)}\equiv D_i N_j + D_j N_i$.
Further, $n^\mu=N^{-1}(1,-N^i)$ is the normal vector to the hypersurface of constant time and $\nabla_\mu$ is the covariant derivative with respect to $g_{\mu\nu}$. Using the notation of \cite{Takahashi:2017pje} one can write the Lagrangian density for the gravitational part (the first term of \eqref{eq:Hamiltonian_analysis_action}) as
\begin{equation}
\mathcal{L}_{\mathrm{grav}} = \sqrt{-g}\,f(\varphi) R = N \sqrt{h} \left[ f\left(\bar{R}+ K_{ij}K^{ij} - K^2\right)-2 K f_{,\varphi} A_{\star} - 2 D_i D^i f\right]\,,
\end{equation}
where we have neglected total derivative terms. $K=K_i^i$ is the trace of the extrinsic curvature, $D_i$ is the covariant derivative with respect to the spatial metric $h_{ij}$, $f_{,\varphi}\equiv \partial_\varphi f$ and
\begin{equation}\label{A*}
A_\star=n^\mu \nabla_\mu\varphi=N^{-1}\left(\dot{\varphi}-N^i \partial_i\varphi\right)\,.
\end{equation}
It is convenient to treat $A_\star$ as an independent variable, by adding to the Lagrangian the condition (\ref{A*}) as a constraint, via a Lagrange multiplier term. Thus, using
\begin{equation}
- g^{\mu\nu}\nabla_\mu \varphi \nabla_\nu \varphi= n^\mu \nabla_\mu \varphi\, n^\nu \nabla_\nu \varphi - h^{\mu \nu} \nabla_\mu \varphi \nabla_\nu \varphi = A_\star^2 - D_i\varphi D^i\varphi\,.
\end{equation}
one can express the mimetic field part of the Lagrangian \eqref{eq:Hamiltonian_analysis_action} as follows
\begin{equation}
\mathcal{L}_{\varphi}=N \sqrt{h} \lambda \left( A_\star^2 - D_i\varphi D^i\varphi -1 \right) - N \sqrt{h} V(\varphi) + \mu \sqrt{h} \left(N A_\star + N^i D_i\varphi - \dot{\varphi}\right)\,.
\end{equation}

\subsection{Canonical momenta and constraints}
Upon the ADM decomposition of the Lagrangian we can calculate the canonical momenta. In view of the relation (\ref{Kij}), the canonical momentum of the metric $h_{ij}$ is
\begin{align}
\pi^{ij}= \frac{\delta \mathcal{L}}{\delta \dot{h}_{ij}}=\frac{1}{2 N}\frac{\delta \mathcal{L}}{\delta K_{ij}} = \sqrt{h} f \left( K^{ij} - K h^{ij}\right) - \sqrt{h} f_{,\varphi} A_\star h^{ij}\,.
\end{align}
This equation can be inverted to get the expressions for the extrinsic curvature and the time derivative of the metric $h_{ij}$ in terms of the momenta $\pi_{ij}$
\begin{align}
K^{ij}&= \frac{1}{\sqrt{h}f} \left(\pi^{ij}-\frac{1}{2}\pi h^{ij}\right) - \frac{f_{,\varphi}}{2 f} A_\star h^{ij}\,, \\
\dot{h}^{ij}&= \frac{N}{\sqrt{h}f}\left(2 \pi^{ij} - \pi h^{ij}\right) - N \frac{f_{,\varphi}}{f} A_\star h^{ij} + 2 D^{(i}N^{j)}\,,
\end{align}
where $\pi=\pi^{i}_i$ is the trace of the canonical momentum. These can be used to get the canonical Hamiltonian density of the gravitational part of the theory
\begin{align}\label{Hgrav}
\mathcal H_{\mathrm{grav}}&=\dot{h}^{ij}\pi_{ij} - \mathcal{L}_{\mathrm{grav}} \nonumber\\
&= N \left[\frac{1}{\sqrt{h}f}\left(\pi^{ij}\pi_{ij}-\frac{1}{2} \pi^2\right) -\sqrt{h} f \bar{R}- A_\star \frac{f_{,\varphi}}{f}\pi - \sqrt{h}\frac{3 f_{,\varphi}^2}{2 f} A_\star^2 + 2 \sqrt{h} D_i D^i f \right] - 2 N^i D_j \pi^j_i\,.
\end{align}
On the right-hand side of (\ref{Hgrav}) we have performed a partial integration with respect to the covariant derivative of the momentum in the last term. Now, the canonical momenta of all other fields are derived straightforwardly, resulting in the following primary constraints
\begin{align}
\label{primary1}
p_\varphi &= \frac{\delta \mathcal{L}}{\delta \dot{\varphi}} = - \sqrt{h}\mu\,, \quad \to \quad \bar{p}_\varphi=p_\varphi + \sqrt{h}\mu \approx 0\,, \qquad & p_\star = \frac{\delta \mathcal{L}}{\delta \dot{A}_\star} \approx 0\,, \\\label{primary 2}
p_\lambda &= \frac{\delta \mathcal{L}}{\delta \dot{\lambda}} \approx 0\,, \qquad & p_\mu=\frac{\delta \mathcal{L}}{\delta \dot{\mu}} \approx 0 \,,\\
\label{primary3}
\pi_N &=\frac{\delta \mathcal{L}}{\delta \dot{N}} \approx 0\,, \qquad &
\pi^i=\frac{\delta \mathcal{L}}{\delta \dot{N}_i} \approx 0\,,
\end{align}
where $\approx$ denotes weak equalities which are only valid on the constraint surface of the phase-space \cite{Dirac:1958sq}.

The total Hamiltonian of the theory including the primary constraints has the following form 
\begin{equation}
H_T = \int \md^3x\,\left( N \mathcal{H} + N^i \mathcal{H}_i + u_\lambda p_{\lambda} + u_\star p_\star + u_\varphi \bar{p}_\varphi + u_\mu p_\mu + u_N \pi_N + u^i \pi_i \right)\,,
\end{equation}
where the so-called Hamiltonian $\mathcal{H}$ and the momentum $\mathcal{H}_i$ constraint are
\begin{align}
\mathcal{H} &= \frac{1}{\sqrt{h}f}\left(\pi^{ij}\pi_{ij}-\frac{1}{2} \pi^2\right) -\sqrt{h} f \bar{R}- A_\star \frac{f_{,\varphi}}{f}\pi - \sqrt{h}\frac{3 f_{,\varphi}^2}{2 f} A_\star^2 + 2 \sqrt{h} D_i D^i f \nonumber \\ & \quad - \sqrt{h} \lambda \left(A_\star^2 - D_i \varphi D^i \varphi -1 \right) + \sqrt{h} V(\varphi) + p_\varphi A_\star\,, \label{eq:Hamiltonian_constraint_mimetic_gravity} \\
\mathcal{H}_i &= - 2 D_j \pi^j_i + p_\varphi D_i \varphi + p_\star D_i A_\star + p_\lambda D_i \lambda + p_\mu D_i \mu\,.
\label{eq:Momentum_constraint_mimetic_gravity}
\end{align}
Note that the last three terms in $\mathcal{H}_i$ vanish separately on the constraint surface \eqref{primary1}-\eqref{primary3}, but we keep them in this form since it elucidates their role as generators of the spatial transformations of the corresponding fields.

As the next step, as usual, we should require the time conservation of the primary constraints. The time conservation of $\pi_N$ and $\pi_i$ yields 
\begin{align}\label{Hconstraints}
\dot{\pi}_N &=\{\pi_N, H_T \} = \mathcal{H}\approx 0\,, \qquad & \dot{\pi}_i = \{\pi_i,H_T \} = \mathcal{H}_i \approx 0\,,
\end{align}
as the conventional secondary constraints inherent to diffeomorphism-invariant gravity theories. 
The time conservation of $p_\lambda$ and $p_\star$ leads to two additional secondary constraints
\begin{align}\label{Cl}
\dot{p}_\lambda&=\{p_\lambda, H_T \} = N \sqrt{h}\left(A_\star^2 - D_i\varphi D^i \varphi - 1 \right) \equiv N C_\lambda \approx 0\,, \\ \label{C*}
\dot{p}_\star&=\{p_\star, H_T \} = N \left(\frac{f_{,\varphi}}{f} \pi + 3 \sqrt{h} \frac{f_{,\varphi}^2}{f} A_\star + 2 \sqrt{h} \lambda A_\star - p_\varphi \right)\equiv N C_\star \approx 0\,,
\end{align}
while the time conservation of $p_\mu$ and $\bar{p}_\varphi$ fixes the values of Lagrange multipliers $u_\varphi$ and $u_\mu$ in terms of other fields and does not give rise to the secondary constraints.
The time conservation of the constraints $C_\lambda$ and $C_\star$ fixes the values of the Lagrange multipliers $u_\star$ and $u_\lambda$, and also does not produce new constraints.

Introducing the smeared Hamiltonian and momentum constraint
\begin{align}
H[\xi] &\equiv \int \md^3x\, \xi\mathcal{H}\,, \\
D[\xi^i] &\equiv \int \md^3x\, \xi^i \mathcal{H}_i\,,
\end{align}
(where $\xi(x)$ and $\xi^i(x)$ are generic functions) one can straightforwardly check that they generate the usual hypersurface deformation algebra of General Relativity
\begin{align}\label{Halgebra}
    \{D[\xi^i],D[\zeta^j]\} &= D[\mathcal{L}_{\xi^i}\zeta^j]\,, \nonumber\\ \{D[\xi^i],H[\xi]\}&=H[\mathcal{L}_{\xi^i}\xi]\,, \\ \{H[\xi],H[\zeta]\}&\approx D\left[h^{ij}\left(\xi \partial_i \zeta - \zeta \partial_i \xi \right)\right]\,,\nonumber
\end{align}
where $\mathcal L_{\xi^i}$ is the Lie derivative along the vector field $\xi^i$.

All in all, the constraints \eqref{primary1}-\eqref{primary3}, \eqref{Hconstraints}, \eqref{Cl} and \eqref{C*} form the full set of the constraints of the system and split into the first and second-class ones, as follows. The eight constraints
\begin{equation}
\pi_{N}, \quad {\mathcal H}, \quad \pi_i, \quad {\mathcal H}_i
\end{equation}
are of the first class, and the six constraints
\begin{equation}
\bar p_\varphi,\quad p_*,\quad p_\lambda, \quad p_\mu, \quad C_\lambda, \quad C_*
\end{equation}
are of the second class. Note that the constraints $p_\lambda$ and $C_\lambda$ are associated with the mimetic matter constraint in the Lagrangian. So, as expected in the formulation under consideration, one does not have an extra first class constraint which would appear in the Hamiltonian analysis of the Weyl-invariant formulation of mimetic gravity, but the two second-class constraints.

The second-class constraints can be regarded to be satisfied in the strong sense, i.e., they can be solved to express certain phase-space variables in terms of others. For this to be consistent with the rules of the Hamiltonian analysis one should pass from the Poisson to the Dirac brackets \cite{Dirac:1958sq}. 

Let us briefly discuss the structure of the Dirac brackets. The Dirac brackets between two phase-space functions $A$ and $B$ are defined as
\begin{align}
\{A,B\}_D = \{A,B\} - \sum\limits_{I,J} \{A,C_I\} \left(\Omega^{-1}\right)^{IJ} \{C_J,B\}\,,
\end{align}
where $C_I$ are the second-class constraints and
\be\label{Omega}
\Omega_{IJ}=\{C_I,C_J\}\,.
\ee
Note that the Dirac brackets of the second-class constraints, with any function of the dynamical variables, are identically zero. Hence the second-class constraints which have been used to construct the Dirac brackets are effectively strongly zero.

Let us first make strongly zero the pair of the constraints $(\bar p_\varphi,p_\mu)$. Their Poisson bracket is 
\be\label{O1}
\Omega=\{\bar p_\varphi,p_\mu\}=\sqrt {h}\,,
\ee
and $p_\mu$ Poisson-commutes with all the other constraints and dynamical variables, except for $\mu$. Therefore, the Dirac brackets constructed with \eqref{O1} of the phase-space functions which do not depend on $(\bar p_\varphi,p_\mu)$ are equal to their Poisson brackets, and we can strongly set $p_\mu=0$ and $\mu=-\frac 1{\sqrt{h}}p_\varphi$.

Thus, we are left with the four second-class constraints
$ p_\lambda,p_\star,C_\lambda$ and $C_\star$.
The corresponding matrix \eqref{Omega} has the following schematic form
\begin{align}
\Omega_{IJ} = 
\begin{pmatrix}
0 & 0 & 0 & A \\
0 & 0 & B & C \\
0 & -B & 0 & D \\
-A & -C & -D & 0 
\end{pmatrix}
\end{align} 
and its inverse is
\begin{align}
\left(\Omega^{-1} \right)^{IJ}=
\begin{pmatrix}
0 & - \frac{D}{AB} & \frac{C}{A B} & -\frac{1}{A} \\
\frac{D}{A B} & 0 & -\frac{1}{B} & 0 \\
-\frac{C}{A B} & \frac{1}{B} & 0 & 0 \\
\frac{1}{A} & 0 & 0 & 0
\end{pmatrix}.
\end{align}
Computing the Poisson brackets between the phase-space variables $\varphi,p_\varphi,h^{ij},\pi^{ij}$ and the constraints $C_J$
\begin{align}
\{p_\varphi,C_J\} = \left(0,0, \star, \star \right)\,, \quad \{h^{ij},C_J\}=(0,0,0,\star)\,, \nonumber \\ \{\varphi,C_J\}=(0,0,0,\star)\,, \quad \{\pi^{ij},C_J\}=(0,0,\star,\star)\,, \nonumber
\end{align}
where $\star$ stands for any non weakly vanishing function, we find that the Dirac brackets between the phase-space variables $\varphi,p_\varphi,h^{ij},\pi^{ij}$ coincide with their Poisson brackets $\{\,,\,\}_D = \{\,,\,\}$.
Hence, we can safely put all the second-class constraints to be strongly zero, without modifying the commutation properties of the rest of the dynamical variables. 

Having identified the number and the nature of the Hamiltonian constraints, we are now in a position to calculate the number of physical degrees of freedom in our model. 
We have $2 \times 14 = 28$ canonical variables (2 x 10 =20 associated with the gravitational field and $2 \times 4=8$ associated with the mimetic scalar $\varphi$, the auxiliary field $A_*$ and the Lagrange multipliers $\mu$ and $\lambda$).
The 8 first-class constraints remove 16 canonical variables, and the 6 second-class constraints remove other 6. So we are left with 6 Hamiltonian degrees of freedom, or 3 Lagrangian degrees of freedom two of which are physical modes of the gravitational field and one is the mimetic scalar mode. This is in agreement with the results of \cite{Chaichian:2014qba,Takahashi:2017pje}.

\subsection{Stability analysis}

\subsection*{Hamiltonian without the mimetic constraint}
It is instructive to first look at the form of the Hamiltonian without the mimetic constraint.
In this case we do not have the canonical variables $\lambda$ and $p_\lambda$ and consequently the two second-class constraints $p_\lambda$ and $C_\lambda$ are absent. The second-class constraints are $p_\star\approx 0$ and $C_\star \approx 0$. We can consider them satisfied in the strong sense and solve $C_\star=0$ for $A_\star$
\begin{equation}
A_\star = \left(p_\varphi - \frac{f_{,\varphi}}{f} \pi \right) \frac{1}{3 \sqrt{h}} \frac{f}{f_{,\varphi}^2}\,.
\end{equation}
Inserting this back into the Hamiltonian we get
\begin{align}
\mathcal{H} =& \frac{1}{\sqrt{h}f} \left(\pi^{ij}\pi_{ij}-\frac{1}{2} \pi^2\right) -\sqrt{h} f \bar{R} + 2 \sqrt{h} D_i D^i f + \sqrt{h} V(\varphi) \nonumber \\
& + \frac{1}{\sqrt{h}}\frac{1}{6f}\left(\pi - \frac{f}{f_{,\varphi}}p_\varphi\right)^2.
\label{eq:Hamiltonian_non_mimetic_constraint}
\end{align}
Due to the time-reparametrization invariance the Hamiltonian is weakly zero. So to describe non-trivial field configurations the different contributions to the Hamiltonian cannot be all positive definite. However, for a stable theory each term should be bounded from below which requires that $f>0$, and $V(\varphi)$ and $D_iD^i f$ are bounded from below, otherwise, for instance, the term $D_iD^i f$ might be a source of gradient instabilities \footnote{Note that the stability criteria here are actually similar to those dynamical systems without reparametrization invariance for which the total Hamiltonian is conserved, i.e. constant in time. Namely, as it is done in the Ostrogradski ghost analysis, one checks whether or not each of the different contributions to the Hamiltonian (whose sum is constant)  may tend to $\pm\infty$.}.

Note that the first and the second term in \eqref{eq:Hamiltonian_non_mimetic_constraint} are not positive definite. So, even if the above constraints on $f$ and $V(\varphi)$ are satisfied, one should still check that for the classical solutions the Hamiltonian does not exhibit instabilities. 
The positive energy theorem for General Relativity, $f=1/2,\;V=0$, indeed states that for asymptotically flat space-times the total energy (ADM-mass) is positive (or zero for a flat space-like hypersurface) as long as the energy-momentum tensor fulfills the dominant energy condition \cite{Witten:1981mf,Schon:1979rg}. 

\subsection*{Hamiltonian with the mimetic constraint}
Now, let us consider the theory with the mimetic constraint. Assuming that the second-class constraints $p_\star$, $C_\star$, $p_\lambda$ and $C_\lambda$ are satisfied in the strong sense one solves the constraint $C_\lambda$ for $A_\star$ and $C_\star$ for $\lambda$ getting
\begin{align}
A_\star &= \pm \sqrt{D_i\varphi D^i \varphi + 1}\,, \label{A*1}\\
\lambda &= \frac{1}{2 A_\star} \frac{p_\varphi}{\sqrt{h}} - \frac{3}{2} \frac{f_{,\varphi}^2}{f} - \frac{1}{2 A_\star}\frac{f_{,\varphi}}{ f} \frac{\pi}{\sqrt{h}}\,.
\label{eq:Lambda}
\end{align}
Upon inserting these expressions into the Hamiltonian constraint we obtain
\begin{align}
\mathcal{H} =& \frac{1}{\sqrt{h}f} \left(\pi^{ij}\pi_{ij}-\frac{1}{2} \pi^2\right) -\sqrt{h} f \bar{R} + 2 \sqrt{h} D_i D^i f + \sqrt{h} V(\varphi) \nonumber \\
&+ \left( p_\varphi -\frac{f_{,\varphi}}{f} \pi - 3 \frac{f_{,\varphi}^2}{f} \sqrt{h} \sqrt{D_i\varphi D^i \varphi + 1} \right) \sqrt{D_i\varphi D^i \varphi + 1} + \frac{3}{2} \frac{f_{,\varphi}^2}{f} \sqrt{h} \left(D_i\varphi D^i \varphi + 1\right)\,,
\label{eq:Hamiltonian_coupled_field_mimetic}
\end{align}
where the positive solution for $A_\star$ was used. The choice of the negative sign in \eqref{A*1} would lead to the same conclusion about the positive definiteness of the mimetic matter contribution into the Hamiltonian \cite{Chaichian:2014qba}.

Comparing the Hamiltonian \eqref{eq:Hamiltonian_coupled_field_mimetic} with that of the non-mimetic model \eqref{eq:Hamiltonian_non_mimetic_constraint} we see that the first four terms are the same, but, instead of the quadratic term with the difference of the two momenta, now there is the 
 term $\frac{3}{2} \frac{f_{,\varphi}^2}{f} \sqrt{h} \left(D_i\varphi D^i \varphi + 1\right)$ which is positive if $f>0$ and
there are terms linear in the momenta $p_\varphi$ and $\pi$.

The term linear in the momentum $p_\varphi$ is expected since it yields the equation of motion of $\varphi$ subject to the mimetic constraint.
The presence of this term may generate a ghost instability since the Hamiltonian is not necessarily bounded from below. There is an instability if the term linear in the momenta can evolve from positive values to negative ones and eventually reach minus infinity. 

Using the equations \eqref{A*1} and \eqref{eq:Lambda} we see that the term in question is of the form $2 \lambda \sqrt{h} \sqrt{D_i\varphi D^i \varphi + 1}$. Therefore, for the mimetic part of the Hamiltonian to be bounded from below $\lambda$ should be non-negative. This condition has already been obtained for the original mimetic dark matter model (with $f=\frac 12$) in \cite{Barvinsky:2013mea,Chaichian:2014qba}. Here we have shown that it is valid for a more general mimetic scalar-tensor theory as well.

\subsection*{Discussion}
The choice of the condition $\lambda>0$ can be understood by directly looking at the action. If $\lambda$ were negative, the field $\varphi$ would have the wrong (ghost-like) sign of its kinetic term.

In order to better understand the dynamics of $\lambda$ and see whether it may change sign with time evolution, let us consider the case of the mimetic dark matter with $f=1/2$ and $V=0$ in \eqref{eq:Hamiltonian_analysis_action}. In this case $\lambda$ is interpreted as the dark matter density, and its negative value would be unphysical. 
In \cite{Chaichian:2014qba} it was argued that, even for the simple mimetic dark matter model, the initial condition $\lambda>0$ might not be preserved in time, since there may exist solutions such that $\lambda$ evolves from positive to negative values and eventually reaches minus infinity. This would mean that the theory becomes unstable. 

Let us elaborate on this issue. In the mimetic dark matter model there exists a conserved Noether current associated with shift-symmetry $\varphi \to \varphi +\mathrm{const}$ 
\begin{equation}\label{J}
\partial_\mu J^\mu\equiv \partial_\mu\left(\sqrt{-g} \lambda g^{\mu\nu}\partial_\nu \varphi \right)=0\,.
\end{equation}
If we fix the general coordinate invariance by imposing a so-called unitary gauge $\varphi=t$ and $N^i=0$, in which $N^2=1$, due to the mimetic constraint, the equation \eqref{J} reduces to \be\label{J1}\partial_t(\sqrt{h}\lambda)= 0\,,
\ee
whose solution is
\be\label{l=}
\sqrt{h}\lambda= p_\varphi/2=C(x^i)\,.
\ee
Therefore, the sign of $\lambda$ is fixed by the choice of the value of $C(x^i)$ at the initial Cauchy surface, which should be non-negative by the stability requirement. Equivalently, the time conservation of $p_\varphi$ could be directly obtained from the Hamiltonian equation for $p_\varphi$ in this gauge. Physically it can be interpreted as the conservation of the energy of the dark matter. We conclude that in the unitary gauge the mimetic dark matter contribution is bounded from below iff $\sqrt{h}\lambda=C(x^i)\geq 0$ . This is consistent with the results of \cite{Ali:2015ftw,Ali:2018vmt}, where the authors used the unitary gauge for deriving a positive definite physical Hamiltonian. 

Now, since under the diffeomorphisms $\lambda$ transforms as a scalar $\lambda^\prime(x^\prime)=\lambda(x)$ and in the gauge $\varphi=t$ and $N^i=0$ the sign of $\lambda$ is appropriately fixed and does not change in time, we conclude that $\lambda$ remains positive or negative for any choice of gauge. In other words, once we found the initial conditions for $\lambda$ be positive in the given gauge, these will determine the allowed choice of the initial conditions in the other gauges.

In a generic mimetic gravity model with $f(\varphi)\neq 1/2$ and $V(\varphi)\neq 0$, the shift-symmetry is broken, and there is no conserved Noether current which ensures that the sign of $\lambda$ is fixed in time.  As was argued in \cite{Chaichian:2014qba}, there may exists initial configurations that would make $\lambda$ to evolve to negative values.  In order to avoid instabilities in the generic case, one can, from the beginning, restrict $\lambda$ to have non-negative values by defining it, e.g. as $\lambda = e^{\hat \lambda}$.\footnote{ Another possibility is to replace $\lambda$ with $\lambda^2$ as was considered by \cite{Babichev:2017lrx} who, in addition, promoted $\lambda$ to a dynamical field for obtaining a caustic free completion of pressureless perfect fluid and k-essence models. } 
Note, however, that as in the case of the Hamiltonian \eqref{eq:Hamiltonian_non_mimetic_constraint} of the GR plus conventional matter, the first two terms of \eqref{eq:Hamiltonian_coupled_field_mimetic} are not positive definite. So the condition $\lambda>0$ does not {\it a priori} guarantee the absence of instabilities, if e.g. Positive Energy Theorem of GR does not apply.

\subsection{Presence of external matter}
Indeed, the presence of instabilities in mimetic gravity for linear perturbations around the FLRW background was discussed in \cite{Langlois:2018jdg,Takahashi:2017pje,Zheng:2017qfs,Firouzjahi:2017txv,Ijjas:2016pad,Hirano:2017zox}. In particular, in \cite{Takahashi:2017pje,Langlois:2018jdg} it was argued that the introduction of additional (conventional) matter into mimetic Horndeski-like theories make them unstable around the FLRW background. 

In this respect it is useful to analyze, in the presence of matter, the properties of the full non-linear Hamiltonian associated with the mimetic systems considered in \cite{Takahashi:2017pje,Langlois:2018jdg}. To this end, let us add to the mimetic action \eqref{eq:Hamiltonian_analysis_action} a perfect fluid action \cite{Matarrese:1984zw}
\begin{equation}\label{Sm}
S_m 
= \int \md^4x\,\sqrt{-g}\, P\left(Y\right)
\equiv \int \md^4x\,\sqrt{-g}\, (-1/2 g^{\mu\nu}\partial_\mu\eta \partial_\nu\eta)^\alpha\,,
\end{equation}
where $Y\equiv -1/2 g^{\mu\nu}\partial_\mu\eta \partial_\nu\eta$ and $\alpha=(1+\omega)/(2\omega)$ fixes the equation of state which describes the ratio between the density $\rho$ and the 
pressure $p$ of the perfect fluid, $p=\omega \rho$ . For simplicity, in the following analysis we will set $\alpha=1$, which describes stiff matter with an equation of state $\omega=1$, and later comment on the case of a generic $\alpha$. The contribution to the mimetic Hamiltonian (eqs. \eqref{eq:Hamiltonian_constraint_mimetic_gravity} and \eqref{eq:Momentum_constraint_mimetic_gravity}) of the matter part has the following form
\begin{equation}
H_\eta = \int \md^3x\,\left(N \mathcal{H}_\eta + N_i \mathcal{H}_\eta^i\right) \;,
\end{equation}
where
\begin{align}
\mathcal{H}_\eta &=\frac{1}{2} \frac{p_\eta^2}{\sqrt{h}} - \frac{1}{2}\sqrt{h}h^{ij} \partial_i\eta \partial_j\eta\,, \\
\mathcal{H}_\eta^i &= p_\eta h^{ij} \partial_j\eta\,.
\end{align}
So that the total Hamiltonian and momentum constraint become
\begin{align}
\mathcal{H}_{\mathrm{tot}}&=\mathcal{H} + \mathcal{H}_\eta\,, \\
{\mathcal H}_{\mathrm{tot}}^i &= \mathcal{H}^i + \mathcal{H}_\eta^i\,.
\end{align}
The structure of the constraints $C_\lambda$ and $C_\star$ (eqs. \eqref{Cl} and \eqref{C*}) does not change since the matter is minimally coupled to the metric and not to the field $\varphi$. From the physical point of view, a coupling between the scalar field $\varphi$ and the conventional matter would introduce interactions between the standard-model particles and the cold dark matter, which have not been observed so far and are normally not considered in the literature.

It is straightforward to see that the usual hypersurface deformation algebra \eqref{Halgebra} is still fulfilled. There is now one additional degree of freedom due to the matter fluid as in the case of standard GR. Solving again the second-class constraints one gets the Hamiltonian constraint in the following form
\begin{align}\label{Hmm}
\mathcal{H} =& \frac{1}{\sqrt{h}f} \left(\pi^{ij}\pi_{ij}-\frac{1}{2} \pi^2\right) -\sqrt{h} f \bar{R} + 2 \sqrt{h} D_i D^i f + \sqrt{h} V(\varphi) + \frac{1}{2} \frac{p_\eta^2}{\sqrt{h}} + \frac{1}{2}\sqrt{h}h^{ij} \partial_i\eta \partial_j\eta\nonumber \\ & + \left( p_\varphi -\frac{f_{,\varphi}}{f} \pi - \frac{3}{2} \frac{f_{,\varphi}^2}{f} \sqrt{h} \sqrt{D_i\varphi D^i \varphi + 1} \right) \sqrt{D_i\varphi D^i \varphi + 1}\,.
\end{align}
The restriction $\lambda>0$ (see Eq. \eqref{eq:Lambda}) again ensures that the contribution of the mimetic matter to the Hamiltonian is bounded from below and the presence of matter does not change this property. 

The above analysis can be extended to a more general case with $\alpha\neq 1$. In general, the derivation of an explicit form of the Hamiltonian will be more involved, but if the matter field $\eta$ is not coupled to the mimetic field $\varphi$, the Hamiltonian of the matter part will never change the structure of the mimetic Hamiltonian. We will discuss this more general case for linear perturbations around the FLRW background.

To summarize, the presence of matter (which does not directly couple to the mimetic scalar) does not change the stability requirement $\lambda>0$ of the mimetic-gravity theory. 
On the other hand, instabilities may also arise due to the non-positive definiteness of the first two terms in the Hamiltonian \eqref{Hmm}. This is what happens for linear perturbations around the FLRW background \cite{Takahashi:2017pje,Langlois:2018jdg}. As we will show, in this case, the instability is caused by a growth of mimetic dust overdensity and is thus of a Jeans type.

\section{Linear analysis around the FLRW background}\label{FLRWB}
In this Section we will revisit results of \cite{Takahashi:2017pje,Langlois:2018jdg}, where it was observed (using the unitary gauge $\varphi=t$) that, in the presence of matter, linear perturbations of mimetic gravity models around the FLRW background have an Ostrogradski ghost.

Above we have shown that the mimetic matter contribution to the full Hamiltonian of gravity plus matter \eqref{Hmm} is bounded, for $\lambda>0$. 

However, the contributions from the pure gravity part are not necessarily bounded, and one has to check the on-shell value of the Hamiltonian explicitly. To identify the origin of the instabilities around the FLRW background in the mimetic dark matter model in the presence of matter described by the generic action \eqref{Sm}, we will derive the corresponding second-order action and Hamiltonian without gauge fixing local reparametrization invariance and give the result in terms of gauge-invariant variables. As a comparison, we present the results for GR with the same matter fluid.

For simplicity, we only consider linear scalar perturbations around the FLRW background whose metric has the following form
\begin{align}
    \mathrm{d}s^2 = - (1 + 2\Phi) \mathrm{d}t^2 + 2 a \partial_i B \md x^i \md t + a^2 \left((1 - 2 \Psi) \delta_{ij} + 2 \partial_i \partial_j E \right) \md x^i \md x^j\,,
\end{align}
where $a(t)$ is the scale factor and $\Phi(x)$, $\Psi(x)$, $B(x)$ and $E(x)$ are scalar perturbations. We will denote the perturbations of the matter scalar field as $\delta \eta$, while $\eta$ will stand for its background value and similar $\delta\varphi$ and $\varphi$ for the mimetic scalar field.
 
 \subsection{General Relativity in the presence of external matter}
 Before discussing the mimetic matter, it is instructive to consider just GR with the matter fluid described by the action \eqref{Sm}. By using the background equations
 \begin{align}
 & 3 H^2 = \frac{2\alpha-1}{2\alpha} \dot\eta^2 P^\prime \,, \\
     & 3 H^2 + 2 \dot H + P =0\,, \\
     & \frac{\partial}{\partial t} \left(a^3 \dot\eta P^\prime \right)=0\,, \quad \to \quad \partial_t(\dot\eta P')=-3H\dot\eta P'\,,
\end{align}
where $H=\dot a/a$ is the Hubble parameter, we can bring the second-order action into the form
\begin{align}
\label{ActionGR+M}
S&= \int \md^3x\,\md t\,a^3\Big[ 3 \alpha \dot{\mathcal{R}}^2 -\frac{3 \alpha}{(2\alpha-1)a^2} (\partial_i \mathcal{R})^2 
+(\alpha-1)3H^2(\hat\Phi)^2-\hat\Phi\left(6\alpha H \dot{\mathcal{R}} -\frac{2\Delta\tilde B}{a^2}\right)
\Big]\,,
\end{align}
where we have introduced
\begin{align}
    \hat\Phi &= \Phi+ \frac{\dot\Psi}{H} - \frac{\dot \eta P^\prime}{2H} \delta \eta\,, \\
    \mathcal{R}&=\Psi + \frac{H}{\dot\eta} \delta\eta\,, \\
    \tilde B &= 2 \Psi + 2 H a^2 \dot E - 2 H a B\,.
\end{align}
The action \eqref{ActionGR+M} has a gauge symmetry under which the field $\Psi$ gets shifted by an arbitrary function $\Psi\to \Psi+\epsilon(x)$, while the other fields transform in such a way that the variables $\mathcal{R},\;\tilde B,\;\hat{\Phi}$ are gauge invariant. Yet, one more local symmetry shifts the scalar fields $E$ and $B$ as follows $E \to E+b$ and $B\to B+a\dot b$. These symmetries can be used to fix $\Psi=0$ and $E=0$, without loss of generality.

After solving the EOM for $\tilde B$ we obtain 
\begin{align}
    S &= \int \md^3x\,\md t\,a^3\Big[ 3 \alpha \dot{\mathcal{R}}^2 -\frac{3  \alpha}{(2\alpha-1)a^2} (\partial_i \mathcal{R})^2 \Big]\,.
\end{align}
We can observe that this action leads to a positive definite second-order Hamiltonian. However, instead of using the gauge-invariant curvature perturbation $\mathcal{R}$ one could equivalently use another gauge-invariant variable 
\begin{align}
    \mathcal{R} = z(t) u\,,
\end{align}
where $z(t)$ is a time-dependent function. In this case, the second-order action takes the form
\begin{align}\label{gs}
    S =& \int \md^3x\,\md t\,a^3 z^2 \Big[ 3 \alpha \dot u^2 - 3 \alpha \frac{\partial_t (a^3 \dot z)}{a^3 z} u^2 -\frac{3 \alpha}{(2\alpha-1)a^2} (\partial_i u )^2 \Big]\,.
\end{align}
Now, we can observe that, for any function $z(t)$ which fulfills
\begin{align}
\frac{\partial_t(a^3 \dot z)}{a^3 z} < 0\,,
\end{align}
there is a tachyon-like instability. Indeed, the Hamiltonian density obtained from \eqref{gs} has the form
\begin{align}\label{Hu}
    \mathcal{H} = a^3 \Big[ \frac{1}{12 \alpha z^2} \frac{p_u^2}{a^6} + z \frac{\partial_t (a^3 \dot z)}{a^3} u^2 + \frac{3 \alpha}{2\alpha-1} z^2 \frac{(\partial_i u )^2}{a^2} \Big]\,,
\end{align}
with $ p_u = 6 \alpha z^2 \dot u$.

The reason of this ambiguity is related to the fact that the Hamiltonian of a system in a time-dependent background is not a conserved quantity, and even if it is positive definite for one choice of phase-space variables, a time-dependent canonical transformation can make it unbounded from below. Hence, the stability behaviour of the system depends, in general, on the choice of observables (for a more detailed discussion of this issue see \cite{Vikman_talk}. The tachyonic instability which showed up in the Hamiltonian \eqref{Hu} could be interpreted as the usual Jeans instability due to the in-falling matter, which vanishes in the ultraviolet-limit. This is in accordance with a discussion in \cite{Gumrukcuoglu:2016jbh} that the Jeans instability of standard matter can be disguised and can be even seen as an infrared ghost instability, and with the suggestion of \cite{DeFelice:2016ucp} to distinguish the Jeans instability from other (dangerous ghost and gradient) instabilities by looking at the ultraviolet limit. 

In the case of the mimetic matter (``dust") we will show that the Jeans instability is seen as a ghost instability at all scales since there is no scale-dependence (due to vanishing sound speed). 

\subsection{Adding the mimetic matter}
 Now let us consider the linear perturbations of mimetic matter in the FLRW background, in the presence of the same external matter fluid. The background equations take the form
 \begin{align}
     & 3 H^2 = \frac{2\alpha-1}{2\alpha} \dot\eta^2 P^\prime + 2\lambda\,, \\
     & 3 H^2 + 2 \dot H + P =0\,, \\
     &\frac{\partial}{\partial t} \left(a^3 \lambda \right) = \frac{\partial}{\partial t} \left(a^3 \dot\eta P^\prime \right)=0 \quad \to \quad \partial_t(\dot\eta P')=-3H\dot\eta P' \quad \to \quad
     \ddot\eta=-\frac 3{2\alpha-1} H\dot\eta\,,
 \end{align}
and the second-order action for the system under consideration can be written as
 \begin{align} \label{FLRWa}
     S=& \int \md^3x\,\md t\,a^3\Big[-3 \dot\xi^2 +\frac{2\alpha-1}{2} P^\prime \dot\chi^2 - 3 \dot \eta P^\prime \dot \xi  \chi + \frac{(\partial_i \xi)^2}{a^2} - \frac{1}{2} P^\prime \frac{(\partial_i \chi)^2}{a^2} + \frac{\Delta \tilde B}{a^2} \left( 2\dot \xi + \dot \eta P^\prime \chi \right) \nonumber \\ 
& + \hat \Phi^2 \left( -3 H^2 + \frac{2\alpha-1}{2} \dot\eta^2 P^\prime +\lambda \right) + \hat\Phi \left(- 2 \frac{\Delta \xi}{a^2} - 2 H \frac{\Delta\tilde B}{a^2} - 2 \lambda \delta_m +6 H \dot\xi - \dot\eta P^\prime (2\alpha-1) \dot\chi \right)
\Big],
\end{align}
where we have introduced the gauge-invariant variables
\begin{align}
\begin{split}
    \xi =& - \Psi - H \delta\varphi\,, \qquad \delta_m= \frac{\delta\lambda}{\lambda} + 3 H \delta\varphi\,, \qquad \chi = \delta\eta - \dot\eta \delta\varphi\,, \\
    \hat \Phi =& \Phi -\delta\dot\varphi\,, \qquad \tilde B= \delta\varphi + a B - a^2 \dot E\,,
\end{split}
\end{align}
with $\delta_m$ having the physical meaning of dark matter overdensity. 

As for GR and the external matter fluid, in equation \eqref{ActionGR+M} the action has two gauge symmetries which can be used to fix e.g. the unitary gauge $\delta\varphi=0$ and $E=0$ without loss of generality.

The variation of the action with respect to $\delta_m$ implies that $\hat \Phi=0$, while the variation with respect to $\hat\Phi$ expresses $\delta_m$ in terms of other variables
\begin{align}
\label{eq:Delta_m}
\lambda \delta_m = - \frac{\Delta \xi}{a^2} - H \frac{\Delta \tilde B}{a^2} + 3 H \dot\xi - \frac{1}{2} \dot\eta P^\prime (2\alpha-1) \dot\chi\,,
\end{align}
where we have used $\hat\Phi =0$. Therefore, the second line of the action \eqref{FLRWa} can be consistently dropped out and we are left with the action
\begin{align} \label{FLRWaa}
     S=& \int \md^3x\,\md t\,a^3\Big[-3 \dot\xi^2 +\frac{2\alpha-1}{2} P^\prime \dot\chi^2 -3 \dot \eta P^\prime \dot \xi \chi + \frac{(\partial_i \xi)^2}{a^2} - \frac{1}{2} P^\prime \frac{(\partial_i \chi)^2}{a^2} + \frac{\Delta \tilde B}{a^2} \left( 2\dot \xi + \dot \eta P^\prime \chi \right) 
\Big]\,,
\end{align}
whose form coincides with that derived in \cite{Takahashi:2017pje,Langlois:2018jdg} in the unitary gauge, but now it is formulated in terms of the gauge-invariant variables.

 \subsubsection{Pure mimetic matter}\label{pmm}
 Before we derive the Hamiltonian for the general case, it is instructive to first elaborate on the case of mimetic matter without the additional matter fluid considered in \cite{Chamseddine:2014vna}.
 
 In this case, we have one degree of freedom which is, however, not propagating in the sense that its sound speed is zero. Indeed, the action \eqref{FLRWaa} reduces to 
 \begin{align} \label{FLRWa1}
     S=& \int \md^3x\,\md t\,a^3\Big[-3 \dot\xi^2 + 2\frac{\Delta \tilde B}{a^2} \dot \xi + \frac{(\partial_i \xi)^2}{a^2} 
\Big]\,.
\end{align}
The equation of motion of $\tilde B$ is
\begin{equation}\label{dotxi}
\Delta\dot\xi=0\,,
\end{equation}
implying that $\Delta\xi$ is time independent, and the equation of motion of $\xi$ is
\begin{equation}\label{dotB}
\Delta \partial_t(a\tilde B)=3\partial_t(a^3\dot\xi)-a\Delta \xi\,.
\end{equation}
Before doing the Hamiltonian analysis let us integrate by parts the second term in \eqref{FLRWa1}. Then we have
\begin{align} \label{FLRWa2}
     S=& \int \md^3x\,\md t\,a^3\Big[-3 \dot\xi^2 - 2\frac{ \dot{\tilde B}}{a^2} \Delta\xi-2 H \tilde B \frac{\Delta\xi}{a^2} + \frac{(\partial_i \xi)^2}{a^2} 
\Big]\,.
\end{align} 
Upon performing the change of variable
\begin{align}
\label{eq:Variable_Transformation}
    \Sigma = \tilde B + \frac{1}{a} \left( \int a\, \md t \right) \xi = \tilde B + \frac{2}{5 H} \xi\,,
\end{align}
(where, in the second step, we have used the background equations to calculate explicitly the integral ($a \propto t^{2/3}$)) one gets the second order action in the following form
\begin{align}
    S = & \int \md^3x\,\md t\,a^3\Big[-3 \dot\xi^2 - 2 \frac{\Delta \xi}{a^2} \dot \Sigma - 2 H \Sigma \frac{\Delta \xi}{a^2}
\Big]\,.
\end{align}
From this action we get the canonical conjugate momenta 
\begin{align}
\label{eq:Canonical_momenta_no_matter}
    \frac{p_\xi}{a^3} = - 6\dot\xi\,, \qquad p_\Sigma = - 2 a \Delta \xi\,,
\end{align}
where the latter is the primary constraint. Consequently, the canonical Hamiltonian density is given by
\begin{align}
    \mathcal{H} = a^3 \Big[ -\frac{1}{12} \frac{p_\xi^2}{a^6} + 2 H \frac{\Delta \xi}{a^2} \Sigma \Big]\,,
\end{align}
and the secondary constraint is
\begin{align}
    C_\Sigma \equiv \Delta p_\xi \approx 0\,.
\end{align}
Assuming suitable initial conditions for $p_\xi$ we can solve the constraints by setting $p_\xi=0$ and $p_\Sigma=-2 a \Delta\xi$. Then the Hamiltonian reduces to
\begin{align}
    \mathcal{H} = - H p_\Sigma \Sigma\,.
\end{align}
We have thus arrived at an Ostrogradski term, which implies that $\Sigma$ is either a ghost (having a negative kinetic energy but a positive mass-squared term) or a tachyon instability. This can be seen by rewriting the Hamiltonian as
\begin{align}\label{Htg}
    \mathcal{H} = \frac{H}{4} \left[ \left(p_\Sigma -\Sigma \right)^2 - \left(p_\Sigma + \Sigma \right)^2 \right]\,.
\end{align}
We can now perform a canonical transformation and call either the first or the second phase-space variable the new canonical momentum or, respectively, the coordinate
\begin{align}
p = \frac{1}{\sqrt{2}} \left( p_\Sigma \mp \Sigma\, \right)\,, \qquad q= \frac{1}{\sqrt{2}} \left( \Sigma \pm p_\Sigma\right)\,. 
\end{align}
Depending on the choice, the Hamiltonian \eqref{Htg} has a tachyon or a ghost instability. At the free level one cannot distinguish between the two, since they just correspond to the canonical transformation which exchanges the role of the generalized coordinate and the canonical momentum and leads to the same equations of motion. So, to understand the nature of the instability it is necessary to extend the consideration to the interacting level. See, for instance, \cite{Blas:2010hb} for a more detailed discussion about the difference of the ghost and the tachyon instability by considering the one-particle exchange amplitude and \cite{Vikman_talk} for a general discussion of ghost instabilities in a similar context. 
\medskip

Solving the Hamiltonian equations of motion
\begin{align}
    \dot p_\Sigma = H p_\Sigma \quad \rightarrow \quad p_\Sigma = a C_1(x)\,, \qquad 
    \dot \Sigma = - H \Sigma \quad \rightarrow \quad \Sigma = \frac{\tilde C_2(x)}{a}\,,
\end{align}
using the equations \eqref{eq:Delta_m}, \eqref{eq:Variable_Transformation}, \eqref{eq:Canonical_momenta_no_matter} and defining $C_2(x) \equiv 2 \Delta \tilde C_2(x)$, we get the expression for the gauge-invariant matter overdensity $\delta_m$
\begin{align}
      \delta_m = - \frac{ C_1(x)}{6 H^2 a^2} \left( 1 - \frac{ H}{a} \int \frac{\md a}{H} \right) - \frac{ C_2(x)}{6 H a^3} = - \frac{ C_1(x)}{10 H^2 a^2} - \frac{ C_2(x)}{6 H a^3}\,.
\end{align}
These are the usual growing and decaying modes as in GR + dust, as discussed in \cite{Lim:2010yk,Chamseddine:2014vna,Arroja:2015yvd}. The growing mode scales with $\delta_m \sim t^{2/3}$ leading to instabilities in the linear perturbation theory. However, the instability is quite slow. It can be interpreted as the usual Jeans instability due to the infalling matter (dust) fitting to our observed tachyon instability. Further, we can note that the condition for a positive Lagrange parameter, $\lambda + \delta\lambda >0 $ or equivalently $\delta_m > -1 $ (in the unitary gauge), requires the integration functions $C_1$ and $C_2$ to be negative. However, this does not remove the growing modes.

Summarizing, even if the necessary condition of the positive definiteness of the Lagrange multiplier field $\lambda$ is satisfied, in the FLRW background the linear fluctuations of the mimetic matter have a ghost or tachyon instability (depending on the choice of phase-space variables), causing the usual Jeans instability of dust which is unstable at all scales. 

\subsubsection{Mimetic dark matter in the presence of external matter}
By adding the external matter fluid, we get two dynamical degrees of freedom. From the action \eqref{FLRWaa} we get the canonical conjugate momenta of $\xi$ and $\chi$
\begin{align}
    \frac{p_\xi}{a^3}=& - 6\dot\xi + 2 \frac{\Delta\tilde B}{a^2} - 3 \dot\eta P^\prime \chi\,, \\
    \frac{p_\chi}{a^3}=& (2\alpha-1) P^\prime \dot \chi\,,
\end{align}
while the momentum $p_B=0$ is a primary constraint. Then the canonical Hamiltonian has the form
\begin{align}
      \mathcal{H}=a^3 \Big[& -\frac{1}{12}\left( \frac{p_\xi}{a^3} + 3\dot\eta P^\prime \chi \right)^2 + \frac{1}{3} \frac{p_\xi}{a^3}\frac{\Delta\tilde B}{a^2} + \frac{1}{2(2\alpha-1)P^\prime} \frac{p_\chi^2}{a^6} - \frac{1}{3}\left(\frac{\Delta\tilde B}{a^2}\right)^2 \nonumber \\
      & - \frac{(\partial_i \xi)^2}{a^2} + \frac{1}{2}P^\prime \frac{(\partial_i \chi)^2}{a^2} \Big]\,.
\end{align}
Further, the time conservation of the primary constraint yields the secondary constraint
\begin{align}
    C_B\equiv \{p_B, \mathcal{H} \}=& -\frac{2}{3}\frac{\Delta\Delta \tilde B}{a^4} + \frac{1}{3} \frac{\Delta p_\xi}{a^5} \approx 0\,.
    \label{eq:Constraint1}
\end{align}
The constraints $p_B$ and $C_B$ are of the second class.
Upon solving the second-class constraints, one can bring the Hamiltonian to the following form
\begin{align}
\label{eq:Hamiltonian_Ostrogradski}
    \mathcal{H} = a^3 \Big[ - \frac{1}{2} \frac{\dot\eta P^\prime}{a^3} \,{p_\xi} \chi + \frac{1}{2(2\alpha-1)P^\prime} \frac{p_\chi^2}{a^6} - \frac{3}{4}\dot\eta^2 P^{\prime 2} \chi^2 - \frac{(\partial_i \xi)^2}{a^2} +\frac{1}{2}P^\prime \frac{(\partial_i \chi)^2}{a^2} \Big]\,.
\end{align}
In accordance with the results of \cite{Takahashi:2017pje,Langlois:2015skt} there is an Ostrogradski ghost instability due to the first term which is linear in the momentum $p_\xi$.

The Hamiltonian \eqref{eq:Hamiltonian_Ostrogradski} equations of motion are
\begin{align}
    \dot \xi =& -\frac{1}{2} \dot\eta P^\prime \chi\,, \label{xichi}\\
    \dot p_\xi =& -2 a \Delta \xi\,, \\
    \dot \chi =& \frac{1}{(2\alpha-1) P^\prime}\frac{p_\chi}{a^3}\,, \\
    \dot p_\chi=& \frac{1}{2}\dot\eta P^\prime p_\xi + \frac{3}{2} a^3 \dot\eta^2 P^{\prime 2} \chi + a P^\prime \Delta \chi\,.
\end{align}
Using the equations of motion we can express the phase-space variables in terms of $\xi$ and its time derivatives
\begin{align}
\label{eq:chi_general}
    \chi =& - \frac{2}{\dot\eta P^\prime} \dot\xi\,, \\ 
\label{eq:p_chi_general}
    p_\chi =& - \frac{2(2\alpha-1) a^3}{\dot\eta } \ddot\xi - \frac{6 (2\alpha-1) a^3 H}{\dot\eta} \dot\xi\,, \\
\label{eq:p_xi_general}
    p_\xi = 
    & - \frac{4(2\alpha-1) a^3}{\dot\eta^2 P^\prime } \dddot\xi - a^3 \left( \frac{24 H}{\dot\eta^2 P^\prime} (2\alpha-1) + \frac{12 H}{\dot\eta^2 P^\prime} \right) \ddot \xi + \frac{4}{\dot\eta^2 P^\prime} a \Delta \dot\xi\nonumber\\
    &-\left(\frac {18 H^2a^3}{{\dot\eta}^2P'} (2\alpha + 1) -\frac{3(4\alpha-1)a^3}{\alpha}\right)\dot\xi\,.
\end{align}
These can be used to derive the fourth-order differential equation for the curvature perturbation $\xi$
\begin{align}
    &\ddddot \xi + \dddot\xi \left( 12 H + \frac{6 H}{2\alpha-1} \right) - \frac{\Delta \ddot \xi}{a^2(2\alpha-1)} - \frac{4 H (2\alpha-1) + 3 H }{(2\alpha-1)^2} \frac{\Delta \dot\xi}{a^2} - \frac{\dot\eta^2 P^\prime}{2(2\alpha-1)} \frac{\Delta \xi}{a^2} \nonumber\\
    & +\ddot \xi \left( \frac{ 9 ( (3\alpha + 2) (2\alpha-1) + 1) H^2}{(2\alpha-1)^2} - \frac{3 (4\alpha-1) \dot\eta^2 P^\prime}{2 (2\alpha-1) \alpha }\right) + \dot \xi \left( \frac{54 (\alpha+1) H^3}{(2\alpha-1)^2} - \frac{(54 \alpha +9) \dot\eta^2 P^\prime H}{4 (2\alpha-1) \alpha } \right)= 0\,.
    \label{eq:EOM_For_Xi}
\end{align}
The same equation can be obtained directly from the variation of the action \eqref{FLRWaa} and taking into account the constraint \eqref{xichi}.

\subsubsection{Dispersion relation in the UV-limit}
Similar to \cite{Babichev:2018twg} we use the ansatz
\begin{align}
\label{eq:Ansatz}
    \xi(x,t) = \xi_0\, e^{\imath \left( \int \omega\,\md t - k_i x^i \right)}\,,
\end{align}
where $\imath$ is the imaginary unit. We are only considering the ultraviolet-limit (UV-limit) in which $H,\lambda,\dot\eta \ll k$.
Further, we are assuming that $\omega$ evolves very slowly in time and one can approximate the time evolution by $\dot \omega/\omega \sim g(H,\dot\eta,\lambda) \ll k $ in the UV-limit with some arbitrary function $g$ and similar for higher derivatives. Later, we will check that this assumption is indeed valid.
Using the UV-limit we can derive the dispersion relation
\begin{align}
     & \omega^4 - 6 \imath \omega^2 \dot\omega - \imath \omega^3 \frac{6 (4\alpha-1) H}{2\alpha-1} - \Bigg( \frac{k^2 }{a^2(2\alpha-1)} + \frac{ 9 ( (3\alpha + 2) (2\alpha-1) + 1) H^2}{(2\alpha-1)^2} - \frac{3 (4\alpha-1) \dot\eta^2 P^\prime}{2 (2\alpha-1) \alpha }\Bigg) \omega^2 \nonumber \\ & + \imath \left( \frac{k^2}{a^2 (2\alpha-1)} + \frac{ 9 ( (3\alpha + 2) (2\alpha-1) + 1) H^2}{(2\alpha-1)^2} - \frac{3 (4\alpha-1) \dot\eta^2 P^\prime}{2 (2\alpha-1) \alpha } \right) \dot\omega \nonumber \\ & + \imath \left( \frac{ H (8\alpha-1) }{(2\alpha-1)^2} \frac{k^2}{a^2} + \frac{54 (\alpha+1) H^3}{(2\alpha-1)^2} - \frac{(54 \alpha +9) \dot\eta^2 P^\prime H}{4 (2\alpha-1) \alpha } \right) \omega + 
     \frac{\dot\eta^2 P^\prime}{2(2\alpha-1)} \frac{k^2}{a^2} =0\,.
\end{align}
We can see that the dispersion relation in the UV-limit has dependence only on $\omega$ and its first derivative. Using now $\dot \omega= g(H,\dot\eta,\lambda) \omega$ we can solve the dispersion relation in the UV-limit in powers of $k$. Since it is a fourth-order polynomial equation there are four solutions, which split into two propagating modes and two purely damped/growing modes
\begin{align}
    \omega_{1,2} =& \pm \frac{1}{\sqrt{2\alpha-1}}\frac{k}{a} + \imath \left( H \left( 4 + \frac{3}{2 (2\alpha-1)} \right) + \frac{5}{2} g_{1,2} \right) + \mathcal{O}(k^{-1})\,, \\
    \omega_{3,4} =& \imath \frac{g_{3,4}(2\alpha-1) + H (8\alpha-1) \pm \sqrt{\left( g_{3,4}(2\alpha-1) + H (8\alpha-1) \right)^2 - 2 (2\alpha-1)^2 \dot\eta^2 P^\prime }}{2 (2\alpha-1) } + \mathcal{O}(k^{-1})\,.
    \label{eq:Mode2_general}
\end{align}
The unknown function $g$ can be solved iteratively. At $k\to\infty$ for the two propagating ``matter" modes we have $\omega \propto k/a$ and hence $\dot\omega / \omega = g_{1,2}(H,\dot\eta,\lambda)= - H$, which confirms our previous assumption that $\dot\omega/\omega \ll k$ in the UV-limit. It yields 
\begin{align}
\label{eq:Mode1_general}
    \omega_{1,2} =&\pm \frac{1}{\sqrt{2\alpha-1}}\frac{k}{a} + \imath H \frac{3 \alpha }{ (2\alpha-1)} + \mathcal{O}(k^{-1})\,.
\end{align}
The modes are damped ($\alpha>1,\,H(t)>0$) and propagate with the sound speed of the matter fluid $c_m=1/\sqrt{2\alpha-1}$. 

Let us now consider the two non-propagating ``dust" modes. Now the form of $g$ is already relevant at leading order, and so the solution is more involved. The leading order of $k$ is evaluated from the terms of the dispersion relation which are proportional to $k^2/a^2$. For the later discussion, we do not need the exact relation, but we are only interested in the main behaviour. Therefore, let us consider just two specific limits.

At first, let us analyze the case in which the dust dominates over the external matter fluid, i.e. $\lambda \gg \dot\eta^2 P^\prime$. From the dispersion relation we can directly evaluate that one solution is trivial (zero) and another one is $\dot\omega/\omega = g_{3,4}(H,\lambda) = \dot H/H \simeq -3 H/2 $, thus resulting in
\begin{align}
    \omega_{3,4}^{\mathrm{dust}}=& \imath \frac{H (10\alpha+1) \pm H(10\alpha+1)}{4 (2\alpha-1)} + \mathcal{O}(k^{-1})\,.
\end{align}
In regimes in which the dust dominates one of the dust modes is constant while the other one is purely damped. 

As another limit let us now consider the regime in which the external fluid dominates, i.e. $\dot\eta^2 P^\prime \gg \lambda$. In this limit the background equation reduces to $\dot\eta^2 P^\prime \propto H^2$, implying that $\dot\omega/\omega = g_{3,4}(H,\lambda) = \dot H/H \simeq -3\alpha H/(2\alpha-1)$ and hence
\begin{align}
    \omega_{3,4}^{\mathrm{ext}}=& \imath \frac{H (5\alpha - 1) \pm H (\alpha+1)}{4 (2\alpha-1)} + \mathcal{O}(k^{-1})\,.
\end{align}
Now both modes are purely damped. As for the matter modes, $\omega$ slowly evolves in time with $\dot\omega /\omega \propto H \ll k$ in the UV-limit, in accordance with our assumption.

Summarizing, we can conclude that in both limits the two dust modes are non-propagating and are just purely damped or constant. However, even if the curvature perturbation is linearly stable, this does not imply that there are no linear instabilities for all physical observables. One can straightforwardly check that the constant dust mode in the dust domination phase leads to a growing matter overdensity mode $\delta_m$, as in the case without external matter.

As a next step, we should look at the properties of the on-shell Hamiltonian for the different modes independently. Considering only the terms with the highest power of $k$ we obtain the on-shell Hamiltonian for the two matter modes \eqref{eq:Mode1_general} 
\begin{align}
    \mathcal{H}_{\mathrm{on-Shell}}^{\omega_{1,2}}\simeq a^3\Big[- \frac{2}{\dot\eta^2 P^\prime} \frac{(\partial_i \dot\xi)^2}{a^2} + \frac{6(2\alpha-1)}{\dot\eta^2P^\prime} \ddot\xi^2 \Big] = \frac{8 k^4}{a (2\alpha-1) \dot\eta^2 P^\prime} \xi^2\,,
\end{align}
which is positive definite. 
On the other hand, the on-shell Hamiltonian for the two non-propagating dust modes \eqref{eq:Mode2_general} in the UV-limit is given by
\begin{align}
    \mathcal{H}_{\mathrm{on-Shell}}^{\omega_{3,4}}&\simeq
    a^3 \Big[ \frac{4}{\dot\eta^2 P^\prime} \frac{\Delta \dot\xi}{a^2} \dot\xi - \frac{(\partial_i \xi)^2}{a^2} + \frac{2}{\dot\eta^2 P^\prime} \frac{(\partial_i \dot\xi)^2}{a^2} \Big] \simeq - a^3 \Big[- \frac{2}{\dot\eta^2 P^\prime} \frac{(\partial_i \dot \xi)^2}{a^2} - \frac{(\partial_i \xi)^2}{a^2}\Big] \nonumber \\ & \simeq a k^2 \xi^2 \Big[ -1 - \frac{2 ( \imath\, \omega_{3,4})^2}{\dot\eta^2 P^\prime} \Big]\,.
\end{align}
The on-shell Hamiltonian is negative definite, as long as $\omega_{3,4}$ is purely imaginary in the UV-limit, which we have explicitly checked for both the limits of dust and external matter domination. 

Summarizing, there are two damped propagating modes with the usual sound speed of the matter fluid, and there are two purely damped non-propagating modes representing the dust. While the propagating modes have a positive definite on-shell Hamiltonian, the two dust modes have a negative definite ghost-like Hamiltonian. 

At the classical level, there are no linear instabilities for the curvature perturbation. However, for instance, the dust matter overdensity $\delta_m$ has an unstable mode, which grows as a power-law in time for the constant dust mode in the matter domination phase.
Further, the non-propagating ghost modes from the dust can be problematic if we take into account higher-order interaction terms, since, as discussed e.g. in \cite{Carroll:2003st} the (in)stability may depend on the values of the interaction coupling constants. Such an analysis is beyond the scope of this paper.

We have thus elaborated on previous results of \cite{Takahashi:2017pje,Langlois:2018jdg} and have found that the presence of matter does not revive dynamical ghost modes (at least in the UV limit). The ghost modes are non-propagating (with zero sound speed) and are associated with the mimetic matter itself, as in the case in which the conventional scalar fluid is absent. As discussed in Section \ref{pmm} these ghost/tachyon modes cause the usual Jeans instability of dust.

\section{Mimetic gravity with higher-derivative terms}
\label{sec:Higher_Derivative_terms}

The observed constraints on the speed of gravitational waves have banned the presence of any higher-derivative term in the theory except for $F(\Box\varphi)$ terms \cite{Ganz:2018vzg}.  Therefore, as outlined in section \ref{sec:Equivalence_approaches} we would also like to study the stability properties of mimetic gravity models containing this type of terms. 

The Hamiltonian analysis of mimetic gravity with a generic term $F(\Box \varphi)$ in the action has been carried out in \cite{Kluson:2017iem,Zheng:2018cuc}. However, since it is only possible to write down the Hamiltonian implicitly in terms of a general inverse function of $F$, it is quite involved to analyze its stability properties. So in what follows we will restrict our consideration to the stability analysis of the case $F(\Box \varphi)=(\Box \varphi)^2$ described by the action \eqref{Sbox}. Details of the calculations are given in the appendix \ref{sec:Appendix_higher_derivatives} and we mention here just the main results.

The action \eqref{Sbox} can be recast into an equivalent second-order form by introducing two scalar fields $\epsilon(x)$ and $\chi(x)$ \footnote{Alternatively, one could rewrite \eqref{000} by introducing only one auxiliary scalar field instead of two as follows $G(\varphi) (\Box\varphi)^2 \to G(\varphi) ( 2 \chi \Box \varphi - \chi^2 )$. }:
\begin{align}\label{000}
S &=\frac{1}{2} \int \md^4x\,\sqrt{-g}\, R + \int \md^4x\,\sqrt{-g}\, \left( G(\varphi) \chi^2 + \epsilon \left(\chi-\Box \varphi \right) -\lambda \left(g^{\mu\nu}\partial_\mu\varphi\partial_\nu\varphi + 1\right)\right) \nonumber \,,\\
 &=\frac{1}{2} \int \md^4x\,\sqrt{-g}\, R + \int \md^4x\,\sqrt{-g}\, \left( G(\varphi) \chi^2 + \epsilon \chi+ g^{\mu\nu} \partial_\mu \epsilon \partial_\nu \varphi -\lambda \left(g^{\mu\nu}\partial_\mu\varphi\partial_\nu\varphi + 1\right)\right)\,.
\end{align}
The Hamiltonian and the momentum constraint have the following form
\begin{align}\label{Hboxm}
\mathcal{H} =& \mathcal{H}_{\mathrm{gr}} - \lambda \frac{p_\epsilon^2}{\sqrt{h}} -  \frac{p_\epsilon p_\varphi}{\sqrt{h}} + \lambda \left(h^{ij}\partial_i\varphi \partial_j\varphi +1\right) - \sqrt{h} G(\varphi) \chi^2 - \sqrt{h} \epsilon \chi \nonumber \\ &-  \sqrt{h} h^{ij}\partial_i\epsilon \partial_j\varphi\,, \\
\mathcal{H}_i =& \mathcal{H}_{\mathrm{gr},i} + p_\varphi \partial_i\varphi + p_\epsilon \partial_i\epsilon + p_\lambda \partial_i \lambda + p_\chi \partial_i \chi\,,
\end{align}
where $\mathcal{H}_\mathrm{gr}$ and $\mathcal{H}_{\mathrm{gr},i}$ are the usual Hamiltonian and momentum constraint of GR. Together with $\pi_N$ and $\pi_i$ they form the set of eight first class constraints. Further, there are six second-class constraints
\begin{align}
p_\lambda \approx & 0 \,, \\
p_\chi \approx & 0\,, \\ \label{C1}
C_\lambda^{(1)}= & \left( - \sqrt{h} \left(h^{ij}\partial_i\varphi \partial_j\varphi +1\right) + \frac{ p_\epsilon^2}{\sqrt{h}}\right) \approx 0\,, \\
\label{Cchi}
C_\chi = & \sqrt{h} \left(2 G(\varphi) \chi + \epsilon \right) \approx 0\,, \\ 
\label{C2}
 C_\lambda^{(2)} = & \left(- \frac{ p_\epsilon \epsilon}{G(\varphi)} - 2 \frac{p_{\epsilon}}{\sqrt{h}}\partial_i \left(\sqrt{h} h^{ij} \partial_j\varphi \right) +2 \sqrt{h} h^{ij}\partial_i\varphi \partial_j \left(\frac{p_\epsilon}{\sqrt{h}}\right) + 4 \pi^{ij} \partial_i\varphi \partial_j\varphi + 2 \pi \right) \approx 0\,, \\
\label{C3}
 C_\lambda^{(3)}= &   \frac{p_\epsilon  (p_\varphi+2\lambda p_\epsilon)}{\sqrt{h}}\left( \frac{1}{G(\varphi)} - 3  \right) + \epsilon \chi \sqrt{h} \left( 3 - \frac{1}{G(\varphi)  } \right)+ \frac{1}{2} \sqrt{h} \bar{R}    - \sqrt{h} \epsilon \frac{G^\prime(\varphi)}{G^2(\varphi)} \nonumber \\ & +3 \sqrt{h} \chi^2 G(\varphi) -3\frac{\pi^2}{\sqrt{h}} + 6 \frac{\pi^{ij}\pi_{ij}}{\sqrt{h}}+V (\partial_i\varphi,h_{ij},\pi_{ij},\lambda,p_\varphi,\varphi,\epsilon,p_\epsilon,\chi),
\end{align}
where in the last constraint we have collected in $V (\partial_i\varphi,h_{ij},\pi_{ij},\lambda,p_\varphi,\varphi,\epsilon,p_\epsilon,\chi)$ all the terms depending on the spatial derivatives of $\varphi$, whose explicit form is given in \eqref{A13}.
Therefore, the model has three degrees of freedom one of which is that of the mimetic field $\varphi$. We see that the higher-derivative term does not introduce an extra degree of freedom, as was shown in \cite{Takahashi:2017pje,Kluson:2017iem,Zheng:2018cuc}. 
\footnote{A comment here is in order. As one can see from eq. \eqref{C3}, the model has a singular point $G(\varphi)=\frac 13$ (observed and discussed in detail e.g. in \cite{Zheng:2018cuc}), in which the first two terms vanish. This results in the fact that in homogeneous backgrounds in which $\varphi$ is identified with the time flow ($\varphi=t$) the number of the degrees of freedom in this model reduces from three to two. We will not elaborate on this issue here and assume that $G(\varphi)\not =\frac 13$.}
The question is whether the mimetic constraint can cure the instability of the higher-derivative theory. So, before discussing the stability conditions for the mimetic theory it is useful to have a look at the original theory without the mimetic constraint. 

\subsection*{Hamiltonian without the mimetic constraint}
In this case we have eight first-class constraints and two second-class ones
\begin{align}
p_\chi \approx 0\,, && C_\chi = \sqrt{h} \left(2 G(\varphi) \chi +  \epsilon \right) \approx 0\,.
\end{align}
Therefore, we now have four degrees of freedom due to the higher-derivative term. The extra degree of freedom is expected to be an Ostrogradski ghost. Indeed, after solving the second-class constraints we get the following Hamiltonian 
\begin{equation}
\label{eq:Hamiltonian_Higher_Derivative_No_Mimetic}
\mathcal{H} = \mathcal{H}_{\mathrm{gr}} + \frac{1}{4} \sqrt{h} \frac{\epsilon^2}{G(\varphi)} - \frac{p_\epsilon p_\varphi}{\sqrt{h}}-  h^{ij} D_i \varphi D_j \epsilon\,.
\end{equation}
In this Hamiltonian, in general, the last two terms are not bounded from below and may cause ghost and gradient instabilities.

\subsection*{Hamiltonian with the mimetic constraint}
Now, solving the second-class constraints \eqref{C1}-\eqref{C2}  one observes that  the following conditions remove $\epsilon$ and $p_\epsilon$ as independent phase-space variables
\begin{align}\label{peps}
p_\epsilon &= \pm \sqrt{h} \sqrt{h^{ij}\partial_i\varphi \partial_j \varphi +1 }\,, \\
\label{chi}
\chi&=-\frac \epsilon{2G(\varphi)}\,,\\
\label{eps}
\epsilon &=\pm \frac{2 G(\varphi)}{\sqrt{h}\sqrt{h^{ij}\partial_i\varphi \partial_j \varphi +1 } } \left(\pm \sqrt{h} h^{ij}\partial_i\varphi \partial_j \sqrt{h^{ij}\partial_i\varphi \partial_j \varphi +1} + 2 \pi^{ij}\partial_i \varphi \partial_j \varphi + \pi \right) \nonumber \\
& \quad - 2 \frac{G(\varphi)}{\sqrt{h}} \partial_i \left(\sqrt{h} h^{ij}\partial_j\varphi \right)\,,
\end{align}
while the constraint \eqref{C3} produces (with the use of the above expressions and choosing there, for convenience, the negative sign) the relation between $\lambda$ amd $p_\varphi$ 
\begin{align}
  2\sqrt{h} \lambda = & p_\varphi + \frac{G(\varphi)}{3G(\varphi)- 1} \Big[ \frac{1}{2}\sqrt{h} \bar{R} - \frac{\pi^2}{\sqrt{h}} (3 G(\varphi) +1) + 6 \frac{\pi^{ij}\pi_{ij}}{\sqrt{h}} +2 \frac{G^\prime(\varphi)}{G(\varphi)}\pi \Big] \nonumber \\
  & + \tilde V(\partial_i\varphi,h_{ij},\pi_{ij},\lambda,p_\varphi,\varphi)\,,  \label{pvarphi}
\end{align}
where we again collected all the terms containing the derivatives of $\varphi$ into $\tilde V(\partial_i\varphi,h_{ij},\pi_{ij},\lambda,p_\varphi,\varphi)$. 
Now inserting the above expressions into the Hamiltonian \eqref{Hboxm} we get
\begin{align}\label{Hbox}
\mathcal{H} =& \mathcal{H}_\mathrm{gr} + \frac{1}{4} \sqrt{h} \frac{\epsilon^2(\pi_{ij},\varphi,h_{ij})}{G(\varphi)} + p_\varphi \sqrt{h^{ij}\partial_i \varphi \partial_j\varphi + 1} \nonumber \\& + \sqrt{h} h^{k l} \partial_k \varphi \partial_l \left[\frac{2 G(\varphi)}{\sqrt{h}\sqrt{h^{ij}\partial_i\varphi \partial_j \varphi +1 } } \left(-\sqrt{h} h^{ij}\partial_i\varphi \partial_j \sqrt{h^{ij}\partial_i\varphi \partial_j \varphi +1} + 2 \pi^{ij}\partial_i \varphi \partial_j \varphi + \pi \right)\right. \nonumber \\ &\left. + \frac{2G(\varphi)}{\sqrt{h}} \partial_i \left(\sqrt{h} h^{ij}\partial_j\varphi \right) \right].
\end{align}
The second term is bounded from below if $G(\varphi)>0$. At the same time, as in the  scalar mimetic models considered in Section III, one observes in \eqref{Hbox} the presence of the characteristic term  linear in $p_\varphi$. However, in contrast to the similar term in eq. \eqref{eq:Hamiltonian_coupled_field_mimetic} which is proportional to $\lambda$ and is positive definite if $\lambda>0$, now we have $p_\varphi$ which is not positive definite, as one can see from the structure of the relation \eqref{pvarphi}. Hence the linear term in the Hamiltonian \eqref{Hbox} is, in general, not bounded from below.

In summary, we conclude that, in general, the mimetic model described by the action \eqref{Sbox} has three degrees of freedom, but may have ghost or gradient instabilities, in agreement with the results of \cite{Langlois:2018jdg,Takahashi:2017pje,Zheng:2017qfs,Firouzjahi:2017txv,Ijjas:2016pad,Hirano:2017zox}, where this issue was discussed using linear perturbations around the FLRW background.

\section{mimetic $f(R)$ gravity}
\label{sec:mimetic_f(R)}
The mimetic $f(R)$ gravity is broadly discussed in the literature \cite{Nojiri:2014zqa,Leon:2014yua,Myrzakulov:2015qaa,Momeni:2015gka,Odintsov:2015cwa,Odintsov:2015wwp,Haghani:2015zga}. Its action can be written in the following form
\begin{equation}
S = \frac{1}{2}\int\md^4x\,\sqrt{-g}\,f(R) -\int \md^4x\,\sqrt{-g}\,\big( \lambda \left(g^{\mu\nu}\partial_\mu\varphi\partial_\nu\varphi+1\right) + V(\varphi)\big) \,.
\end{equation}
There are several ways to derive the Hamiltonian of the $f(R)$ theory, which are equivalent up to canonical transformations (see \cite{Deruelle:2009pu} for a detailed discussion).
Here we rewrite the action as that of a scalar-tensor theory, by introducing two extra scalar fields $\chi(x)$ and $\mu(x)$ 
\begin{equation}
\label{eq:Action_mimetic_f(R)}
S = \frac{1}{2}\int\md^4x\,\sqrt{-g}\,\left[f(\chi)+\mu\left(R-\chi \right)\right]-\int \md^4x\,\sqrt{-g}\,\big(\lambda \left(g^{\mu\nu}\partial_\mu\varphi\partial_\nu\varphi+1\right) + V(\varphi)\big)\,.
\end{equation}
After the usual foliation of spacetime one gets the following Hamiltonian and momentum constraint (see appendix \ref{sec:Appendix_f_R} for details) 
$$\mathcal{H} = \mathcal{H}_{\mathrm{grav}} +\mathcal{H}_\varphi\,,$$
where
\begin{equation}
\label{eq:H_grav_f_R}
\mathcal{H}_{\mathrm{grav}}= \frac{2}{\sqrt{h}\mu}\left(\pi^{ij}\pi_{ij}-\frac{1}{2}\pi^2 \right) + \frac{1}{3 \sqrt{h} \mu }\left(\mu p_\mu - \pi \right)^2-\frac{1}{2}\sqrt{h}\mu \bar{R} + \frac{1}{2}\sqrt{h}\mu\chi-\frac{1}{2}\sqrt{h}f(\chi)+ \sqrt{h}D_a D^a \mu\,,
\end{equation}
and
\begin{align}
\mathcal{H}_\varphi &= \frac{p_\varphi^2}{4\sqrt{h}\lambda} + \sqrt{h} \lambda \left(h^{ij}\partial_i\varphi\partial_j\varphi+1\right) + V(\varphi)\,,
\end{align}
and 
\begin{align}
\mathcal{H}^i &= - 2 D_j \pi^{i j} + p_\mu \partial^i \mu + p_\chi \partial^i \chi+ p_\varphi \partial^i \varphi + p_\lambda \partial^i \lambda\,.
\end{align}
We have the usual eight (gravity) first-class constraints $\pi_N$, $\mathcal{H}$, $\pi_i$ and $\mathcal{H}_i$, and four second-class constraints
\begin{align}
  p_{\chi} \approx 0\,, \quad C_\chi=\sqrt{h}\left(\mu + f^\prime(\chi)\right)\approx 0\,,\quad p_\lambda \approx 0\,, \quad C_\lambda= - \frac{p_\varphi^2}{4 \sqrt{h} \lambda^2} + \sqrt{h}\left(h^{ij}\partial_i\varphi\partial_j\varphi+1\right) \approx 0\,,
\end{align}
Therefore, we obtain the expected result that the theory has four degrees of freedoms.

Upon solving the second-class constraints we get the following Hamiltonian
\begin{equation}\label{Hsvt}
\mathcal{H} = \mathcal{H}_{\mathrm{grav}} + \mathcal{H}_\varphi = \mathcal{H}_{\mathrm{grav}} + p_\varphi \sqrt{h^{ij}\partial_i\varphi\partial_j\varphi +1}+ V(\varphi)\,,
\end{equation}
which, as in previous cases, has a term linear in $p_\varphi=2\lambda\sqrt{h}\sqrt{h^{ij}\partial_i\varphi\partial_j\varphi +1}$.
Again, requiring that $\lambda >0$ we ensure that the mimetic scalar field part of the Hamiltonian is bounded from below. 
In addition to this mimetic constraint, one has the usual stability conditions on the $f(R)$ gravity theory, which remain unaltered. 

Note, that the gravity part of the Hamiltonian \eqref{eq:H_grav_f_R} is equivalent to the Hamiltonian of the scalar-tensor theory \eqref{eq:Hamiltonian_non_mimetic_constraint} with the identifications
\begin{align}
    \tilde f(\mu) = \frac{\mu}{2}\,, \qquad \tilde V(\mu) = \frac{1}{2} \mu \chi(\mu) - \frac{1}{2} \mu f(\chi(\mu))\,,
\end{align}
where $\chi(\mu)$ is the formal solution of the second class constraint $C_\chi$ for $\chi$ in terms of the scalar field $\mu$. This confirms the well-known relation between standard $f(R)$ gravity and a scalar-tensor theory. 

From the above consideration one can easily conclude that the obtained results can be generalized to any scalar-vector-tensor theory of the form 
\begin{equation}
S = S(g_{\mu\nu},\chi_1,...,\chi_n, A_1^\mu,...,A_m^\mu) - \int \md^4x\,\sqrt{-g}\Big( \lambda \left(\partial_\mu \varphi \partial^\mu \varphi +1 \right)+V(\varphi)\Big)\,,
\end{equation}
where $\chi_n$ and $A_m^\mu$ are scalar and vector fields which are not directly coupled to the mimetic scalar $\varphi$.

The Hamiltonian of this model has a form similar to \eqref{Hsvt}. Hence, the stability requirements for the initial $f(R)$ theory without the mimetic field remain unaltered by the presence of the latter, 
if $\lambda>0$.

\section{Conclusions}
\label{sec:Conclusions}
In this paper, we have carried out the stability analysis of the full Hamiltonian for several mimetic gravity models. The mimetic contribution to the Hamiltonian of the most general mimetic scalar-tensor theory, restricted to a healthy primary seed action compatible with the constraint that the speed of gravitational waves equals the speed of light \cite{Ganz:2018vzg}, has been shown to be free of any dangerous instability, if the mimetic energy-density field $\lambda$ is positive definite $\lambda > 0$. This is in agreement with the results of \cite{Barvinsky:2013mea,Chaichian:2014qba} for the original mimetic dark matter model, in which case, as we have shown, the shift-symmetry of the mimetic scalar field $\varphi$ ensures that the sign of $\lambda$ is not changed in time. In general, one should {\it a priori} impose the condition $\lambda>0$ into the mimetic action. We have also discussed the role of conventional matter for the stability of the mimetic scalar-tensor theory. Using the example of a fluid, we have shown that the necessary stability condition $\lambda>0$ of the mimetic gravity also persists in the presence of matter, at least if it does not mix with the mimetic sector. 
However, in general, the gravity part of the Hamiltonian is not bounded from below and can lead to instabilities if the Positive Energy Theorem of GR does not apply. 

The same result also applies to mimetic $f(R)$ gravity, which is healthy, if the usual stability conditions of the standard $f(R)$ gravity are assumed and $\lambda>0$.

In contrast, if we add mimetic matter to an unhealthy seed action, the resulting mimetic gravity theory remains, in general, unstable. As an example, we have considered a scalar-tensor theory with a single higher-derivative term $(\Box \varphi)^2$, which contains an Ostrogradski ghost. The addition of the mimetic constraint on $\varphi$ eliminates one degree of freedom, as discussed by \cite{Takahashi:2017pje,Kluson:2017iem}, however, the mimetic theory contains instabilities anyway.

We have also revisited results of \cite{Takahashi:2017pje,Langlois:2018jdg} regarding stability issues of linear perturbations around the FRLW background of the mimetic dark matter in the presence of scalar matter. We have found that the presence of conventional matter does not revive dynamical ghost modes (at least in the UV limit). The modes with non-positive Hamiltonian are non-propagating (with zero sound speed) and are associated with the mimetic matter itself, as in the case in which the conventional scalar fluid is absent. These ghost/tachyon-like modes cause the usual Jeans instability of dust. To trace the fate of this instability one should go to the interaction level, which is beyond the scope of this paper.

\begin{acknowledgments}
We would like to thank I. Bandos, N. Bartolo, M. Chaichian, J. Kluso\u{n}, T. Kobayashi, T. Koivisto, S. Nojiri, I. Oda, M. Oksanen, S. Ramazanov, A. Ricciardone, S. Sibiryakov, A. J. Tolley, A. Tureanu, and especially A. Barvinsky and A. Vikman, for useful discussions and comments on a preliminary version of this paper.
Part of the computations are done using Mathematica\footnote{https://www.wolfram.com/mathematica/} with the algebra package xAct\footnote{http://www.xact.es/} and its contributed package xPand\footnote{http://www2.iap.fr/users/pitrou/xpand.htm}. Work of D.S. was supported in part by the Russian Science Foundation grant 14-42-00047 in association with Lebedev Physical Institute and by the Australian Research Council project No. DP160103633. S.M. acknowledges partial financial support by ASI Grant No. 2016- 24-H.0. P.K. acknowledges financial support from ``Fondazione Ing. Aldo Gini''. 

\end{acknowledgments}

\appendix

\section{Calculations for mimetic gravity with higher-derivative terms}
\label{sec:Appendix_higher_derivatives}

The starting action is
\begin{align}
S =\frac{1}{2} \int \md^4x\,\sqrt{-g}\, R + \int \md^4x\,\sqrt{-g}\, \left( G(\varphi) (\Box \varphi)^2 -\lambda \left(g^{\mu\nu}\partial_\mu\varphi\partial_\nu\varphi + 1\right)\right)\,.
\end{align}
Using the notation from \cite{Kluson:2017iem}, this can be rewritten as
\begin{align}
S &=\frac{1}{2} \int \md^4x\,\sqrt{-g}\, R + \int \md^4x\,\sqrt{-g}\, \left( G(\varphi) \chi^2 + \epsilon \left(\chi-\Box \varphi \right) -\lambda \left(g^{\mu\nu}\partial_\mu\varphi\partial_\nu\varphi + 1\right)\right) \nonumber \\
 &=\frac{1}{2} \int \md^4x\,\sqrt{-g}\, R + \int \md^4x\,\sqrt{-g}\, \left( G(\varphi) \chi^2 + \epsilon \chi+ g^{\mu\nu} \partial_\mu \epsilon \partial_\nu \varphi -\lambda \left(g^{\mu\nu}\partial_\mu\varphi\partial_\nu\varphi + 1\right)\right)\,.
\end{align}
The ADM decomposition yields
\begin{align}
\mathcal{L}=& N \sqrt{h} \Big(\frac{1}{2} \left( K^{ij}K_{ij}-K^2+\bar{R} \right) - \lambda \left(-\nabla_n \varphi \nabla_n \varphi + h^{ij} \partial_i\varphi \partial_j \varphi + 1 \right) + G(\varphi) \chi^2 \nonumber \\ & + \epsilon \chi - \nabla_n \epsilon \nabla_n\varphi + h^{ij} \partial_i\epsilon \partial_j \varphi \Big)\,,
\end{align}
where we used the notation $\nabla_n \varphi = \left(\dot{\varphi}-N^i\partial_i\varphi\right)/N$.
The canonical conjugate momenta are 
\begin{align}
\pi^{ij}&=\frac{\delta \mathcal{L}}{\delta \dot{h}_{ij}} = N \sqrt{h} \left(K^{ij}-h^{ij} K\right)\,, \\
p_\varphi &= \frac{\delta \mathcal{L}}{\delta \dot{\varphi}} = \sqrt{h} \left( 2 \lambda \nabla_n \varphi - \nabla_n\epsilon \right)\,, \\
p_\epsilon &= \frac{\delta \mathcal{L}}{\delta \dot{\epsilon}} = - \sqrt{h} \nabla_n \varphi\,.
\end{align}
The other momenta are primary constraints $\pi_N=\pi^i = p_\lambda=p_\chi=0$. The extended Hamiltonian can be written as
\begin{equation}
H_T = \int \md^3 x \,\left( N \mathcal{H} + N^a \mathcal{H}_a + u_\lambda p_\lambda + u^i \pi_i + u_N \pi_N + u_\chi p_\chi \right)
\end{equation}
with the Hamiltonian and momentum constraint having the following form
\begin{align}
\mathcal{H} =& \mathcal{H}_{\mathrm{gr}} - \lambda \frac{p_\epsilon^2}{\sqrt{h}} -  \frac{p_\epsilon p_\varphi}{\sqrt{h}} +  \lambda \left(h^{ij}\partial_i\varphi \partial_j\varphi +1\right) -  \sqrt{h} G(\varphi) \chi^2 -  \sqrt{h} \epsilon \chi \nonumber \\ &- \sqrt{h} h^{ij}\partial_i\epsilon \partial_j\varphi\,, \\
\mathcal{H}_i =& \mathcal{H}_{\mathrm{gr},i} + p_\varphi \partial_i\varphi + p_\epsilon \partial_i\epsilon + p_\lambda \partial_i \lambda + p_\chi \partial_i \chi\,,
\end{align}
where $\mathcal{H}_\mathrm{gr}$ and $\mathcal{H}_{\mathrm{gr},i}$ are the usual Hamiltonian and momentum constraint from GR.

The time conservation of the primary constraints $\pi_N$ and $\pi_i$ yields the usual Hamiltonian and momentum constraints, while, due to the conservation of $p_\lambda$ and $p_\chi$, one obtains
\begin{align}
\dot{p}_\lambda &= \{p_\lambda, H_T\} = N \left( -  \sqrt{h} \left(h^{ij}\partial_i\varphi \partial_j\varphi +1\right) + \frac{ p_\epsilon^2}{\sqrt{h}}\right) \equiv N C_\lambda^{(1)} \approx 0\,, \\
\dot{p}_\chi &= \{p_\chi,H_T\}= N \sqrt{h} \left(2 G(\varphi) \chi +  \epsilon \right) \equiv N C_\chi \approx 0\,.
\end{align}
The time conservation of the secondary constraint $C_\chi$ fixes the Lagrange parameter $u_\chi$ while the conservation of $C_\lambda^{(1)}$ yields a tertiary constraint
\begin{align}
\dot{C}_\lambda &= \{C_\lambda, H_T\} \approx N \left(- \frac{ p_\epsilon \epsilon}{G(\varphi)} - 2 \frac{p_{\epsilon}}{\sqrt{h}}\partial_i \left(\sqrt{h} h^{ij} \partial_j\varphi \right) +2 \sqrt{h} h^{ij}\partial_i\varphi \partial_j \left(\frac{p_\epsilon}{\sqrt{h}}\right) + 4 \pi^{ij} \partial_i\varphi \partial_j\varphi + 2 \pi \right) \nonumber \\ &\equiv N C_\lambda^{(2)} \approx 0\,,
\end{align}
where the constraints $C_\lambda$ and $C_\chi$ were used. The time conservation of $C_\lambda^{(2)}$ yields a further constraint, namely
\begin{align}\label{A13}
    C_\lambda^{(3)} \equiv & \frac{1}{N} \{C_\lambda^{(2)},H_T\} \nonumber \\ \approx & \lambda \left( - 6 \frac{p_\epsilon^2}{\sqrt{h}} + \frac{2 p_\epsilon^2}{G(\varphi)\sqrt{h}} - 2 \sqrt{h} h^{ij} \partial_i\varphi \partial_j \varphi \right) + \frac{p_\epsilon p_\varphi}{\sqrt{h}} \left( \frac{1}{G(\varphi)} - 3 - 2 D_i \varphi D^i\varphi \right) + \frac{1}{2} \sqrt{h} \bar{R}  \nonumber \\ & + \epsilon \chi \sqrt{h} \left( 3 - \frac{1}{G(\varphi)  } + 2 D_i\varphi D^i \varphi \right) + 4 \frac{p_\epsilon}{\sqrt{h}} D_i D^i p_\epsilon - 4 \sqrt{h} \chi D_i D^i \varphi + 6 \frac{\pi }{\sqrt{h}} D_i \varphi D^i p_\epsilon  \nonumber \\ & - \sqrt{h} \frac{1}{G(\varphi)} D_i \varphi D^i \epsilon + \sqrt{h} D_i\varphi D^i \epsilon   +\sqrt{h}\bar{R} D_i\varphi D^i\varphi  + 4 \frac{\pi^{kl}\pi_{kl}}{\sqrt{h}} D_i \varphi D^i \varphi  - \sqrt{h} \epsilon \frac{G^\prime(\varphi)}{G^2(\varphi)} \nonumber \\ & + \sqrt{h} \chi^2 G(\varphi) ( 3 + 2 D_i\varphi D^i \varphi) - \frac{\pi^2}{2\sqrt{h}} ( 6 + 4 D_i \varphi D^i \varphi) + 8 \frac{p_\epsilon}{\sqrt{h}} D^i \varphi D_j \pi^{j}_i + 8 \frac{p_\epsilon \pi^{ij}}{\sqrt{h}} D_i D_j \varphi \nonumber \\ &  + 2\sqrt{h} \left(D_i D^i \varphi \right)^2 - 4 \sqrt{h} D^i \varphi D_j D^j D_i\varphi - 16 \frac{\pi_i^k\pi_{jk}}{\sqrt{h}} D^i \varphi D^j\varphi -2 \sqrt{h} D_i\varphi D^i \epsilon D_j \varphi D^j\varphi \nonumber \\ & - 2 \frac{p_\epsilon \pi }{\sqrt{h}} D_i D^i \varphi + 6 \frac{\pi^{ij}\pi_{ij}}{\sqrt{h}} - 16 \frac{\pi^{ij}}{\sqrt{h}} D_i p_\epsilon D_j\varphi + 8 \frac{\pi \pi^{ij}}{\sqrt{h}} D_i\varphi D_j\varphi - 2 \sqrt{h} D_i D_j\varphi D^i D^j \varphi, 
\end{align}
where we have used the previous constraints to simplify the expression. The new constraint $C_\lambda^{(3)}$ explicitly depends on $\lambda$ and consequently the time conservation of it fixes the Lagrange parameter $u_\lambda$ for the primary constraint $p_\lambda.$

By a straightforward calculation one can check that the usual hypersurface deformation algebra is fulfilled
\begin{align}
\begin{split}
\{D[\xi^i],D[\zeta^j]\}&=D[\xi^i \partial_i \zeta^j - \zeta^i \partial_i \xi^j]\,, \\ 
\{D[\xi^i],H[\xi]\}&=H[\mathcal{L}_{\xi^i}\xi]\,, \\
\{H[\xi],H[\zeta]\} &\approx D\left[h^{ij}\left(\xi \partial_i \zeta -\zeta \partial_i \xi \right)\right]\,.
\end{split}
\end{align}
 The Dirac matrix $\Omega_{IJ}=\{C_I,C_J\}$, where $C_I$ are the six second class constraints $p_\lambda,\;p_\chi,\;C_\chi,\;C_\lambda^{(1)},\;C_\lambda^{(2)}$ and $C_\lambda^{(3)}$, and its inverse can be expressed as
\begin{align}
   \Omega_{IJ} =
\begin{pmatrix}
0 & 0 & 0 & 0 & 0 & A \\
0 & 0 & B & 0 & 0 & C \\
0 & -B & 0 & D & E & F \\
0 & 0 & -D & 0 & G & H \\
0 & 0 & -E & -G & I & J \\
-A & - C & -F & -H & -J & -K 
\end{pmatrix}, \quad
\left(\Omega^{-1}\right)^{IJ}=
\begin{pmatrix}
\star & \star & \star & \star & \star & \star \\
\star & \star & \star & \star & \star & 0 \\
\star & \star & 0 & 0 & 0 & 0\\
\star & \star & 0 & \frac{I}{G^2} & -\frac{1}{G} & 0 \\
\star & \star  & 0 & \frac{1}{G} & 0 & 0 \\
\star & 0 & 0 & 0 & 0 & 0
\end{pmatrix}. 
\end{align}
By using the Poisson brackets between the remaining phase-space variables and the constraints
\begin{align}
&\{p_\varphi,C_I\}=(0, 0, \star ,\star,\star,\star)\,, \quad \{\varphi,C_I\}=(0,0,0,0,0,\star)\,, \nonumber \\
&\{h^{ij},C_I\}=(0,0,0,0 ,\star, \star )\,,\quad \{\pi^{ij},C_I\}=(0,0,0,\star,\star,\star)\,,
\end{align}
we can see that the structure of the Dirac brackets is in general different from that of the Poisson brackets. Only for the case of the scalar field $\varphi$ the Dirac bracket coincides with the Poisson bracket. Further, for the Dirac brackets only the terms $G=\{C_\lambda^{(1)},C_\lambda^{(2)}\}$ and $I=\{C_\lambda^{(2)},C_\lambda^{(2)}\}$ are relevant.

\section{Calculations for mimetic $f(R)$ gravity}
\label{sec:Appendix_f_R}
The action for mimetic gravity can be written as \eqref{eq:Action_mimetic_f(R)}
\begin{equation}
S = \frac{1}{2}\int\md^4x\,\sqrt{-g}\,\left[f(\chi)+\mu\left(R-\chi \right)\right]-\int \md^4x\,\sqrt{-g}\,\big(\lambda \left(g^{\mu\nu}\partial_\mu\varphi\partial_\nu\varphi+1\right) + V(\varphi)\big)\,.
\end{equation}
Using the standard ADM mechanism one can decompose the Lagrangian for the gravitational part 
\begin{equation}
\mathcal{L}_{\mathrm{grav}} = \frac{1}{2}N\sqrt{h}\left(f(\chi) + \mu\left(K_{ij}K^{ij}-K^2 +\bar{R}-\chi\right)-2K \nabla_n \mu - 2 D_iD^i \mu \right)\,.
\end{equation}
The same can be done for the part of the mimetic constraint
\begin{equation}
\mathcal{L}_\varphi = - N \sqrt{h} \left(\lambda \left(-\frac{\dot{\varphi}^2}{N^2}+ 2 \frac{N^i}{N^2}\dot{\varphi}\partial_i\varphi+ \left(h^{ij}-\frac{N^iN^j}{N^2}\right)\partial_i\varphi\partial_j\varphi + 1\right) + V(\varphi) \right)\,.
\end{equation} 
The canonical conjugate momenta are
\begin{align}
&\pi^{ij} = \frac{\delta \mathcal{L}}{\delta \dot{h}_{ij}}= \frac{1}{2}\sqrt{h}\mu \left(K^{ij}-h^{ij}K\right)-\frac{1}{2}\sqrt{h}h^{ij}\nabla_n\mu\,, \quad p_\mu =\frac{\delta \mathcal{L}}{\delta \dot{\mu}}=-\sqrt{h}K\,, \quad \pi^i=\frac{\delta \mathcal{L}}{\delta \dot{N}^{i}}=0\,, \nonumber \\
& p_\varphi=\frac{\delta \mathcal{L}}{\delta \dot{\varphi}}=\frac{2}{N}\sqrt{h}\left(\dot{\varphi}-N^i\partial_i\varphi\right)\,, \quad p_\chi=\frac{\delta \mathcal{L}}{\delta \dot{\chi}}=0\,, \quad p_\lambda=\frac{\delta \mathcal{L}}{\delta \dot{\lambda}}=0\,, \quad \pi_N=\frac{\delta \mathcal{L}}{\delta \dot{N}}=0\,.
\end{align}
From these we get the expressions for the velocities
\begin{align}
\dot{\mu}=\frac{2N}{3\sqrt{h}}\left(p_\mu \mu - \pi \right) + N^i \partial_i \mu\,, \quad \dot{\varphi}= \frac{N}{2\sqrt{h}\lambda}p_\varphi + N^i \partial_i\varphi\,, \nonumber \\ \dot{h}_{ij}=\frac{N}{\sqrt{h}}\left(4 \frac{\pi_{ij}}{\mu}- \frac{4}{3}h_{ij}\frac{p}{\mu}-\frac{2}{3}h_{ij}p_\mu\right) +2 D_{(i}N_{j)}\,.
\end{align}
The extended Hamiltonian is
\begin{align}
H_T = \int \md^3x\,\left( N \mathcal{H} + N^a \mathcal{H}_a + u_\lambda p_\lambda + u_\chi p_\chi + u_i \pi^i + u_N \pi_N \right)
\end{align}
with $\mathcal{H} = \mathcal{H}_{\mathrm{grav}} +\mathcal{H}_\varphi$, where
\begin{align}
H_{\mathrm{grav}}&= \frac{2}{\sqrt{h}}\left(\frac{\pi^{ij}\pi_{ij}}{\mu}-\frac{1}{3}\frac{\pi^2}{\mu}-\frac{1}{3}\pi p_\mu + \frac{1}{6} \mu\, p_\mu^2\right)-\frac{1}{2}\sqrt{h}\mu \bar{R} + \frac{1}{2}\sqrt{h}\mu\chi-\frac{1}{2}\sqrt{h}f(\chi)+ \sqrt{h}D_a D^a \mu\,, \\
\mathcal{H}_\varphi &= \frac{p_\varphi^2}{4\sqrt{h}\lambda} + \sqrt{h} \lambda \left(h^{ij}\partial_i\varphi\partial_j\varphi+1\right) + V(\varphi)\,,
\end{align}
and
\begin{align}
\mathcal{H}^i = - 2 D_j \pi^{i j} + p_\mu \partial^i \mu + p_\chi \partial^i \chi + p_\varphi \partial^i \varphi + p_\lambda \partial^i \lambda\,.
\end{align}
We obtain the usual Hamiltonian constraint $\mathcal{H}\approx 0$ and the momentum constraint $\mathcal{H}_i\approx 0$, due to the time conservation of $\pi_N$ and $\pi_i$.

The conservation of $p_\lambda$ and $p_\chi$ yields
\begin{align}
\dot{p}_\lambda &= \{p_\lambda,H_T\}= - N \left(-\frac{p_\varphi^2}{4 \sqrt{h}\lambda^2}+\sqrt{h}\left(h^{ij}\partial_i\varphi\partial_j\varphi\right)\right) \equiv - N C_\lambda \approx 0\,, \\
\dot{p}_\chi &= \{p_\chi,H_T\}= - \frac{1}{2} N \sqrt{h} \left(\mu + f^\prime(\chi)\right)\equiv -\frac{1}{2} N C_\chi \approx 0\,.
\end{align}
The time conservation of $C_\lambda$ fixes the Lagrange parameter $u_\lambda$, while the time conservation of $C_\chi$ fixes the Lagrange parameter $u_\chi$ if $f^{\prime\prime}(\chi)\neq 0$, which will be assumed in the following.

It is straightforward to check that the hypersurface deformation algebra is fulfilled
\begin{align}
\begin{split}
\{D[\xi^i],D[\zeta^j]\}&=D[\xi^i \partial_i \zeta^j - \zeta^i \partial_i \xi^j]\,, \\ 
\{D[\xi^i],H[\xi]\}&=H[\mathcal{L}_{\xi^i}\xi]\,, \\
\{H[\xi],H[\zeta]\} &\approx D\left[h^{ij}\left(\xi \partial_i \zeta -\zeta \partial_i \xi \right)\right]\,.
\end{split}
\end{align}

From the four second-class constraints $C_I=\{p_\lambda,p_\chi,C_\lambda,C_\chi\}$ we get the Dirac matrix $\Omega_{IJ}=\{C_I,C_J\}$ and its inverse 
\begin{align}
\Omega_{IJ}&=
\begin{pmatrix}
0 & 0 & A & 0 \\
0 & 0 & 0 & B \\
-A & 0 & D & 0 \\
0 & -B & 0 & 0
\end{pmatrix}, \quad
\left(\Omega^{-1}\right)^{IJ}=
\begin{pmatrix}
\frac{D}{A^2} & 0 & -\frac{1}{A} & 0 \\
0 & 0 & 0 & -\frac{1}{B} \\
\frac{1}{A} & 0 & 0 & 0 \\
0 & \frac{1}{B}& 0 & 0
\end{pmatrix}.
\end{align}
Since the remaining phase-space variables $\varphi,p_\varphi,h^{ij},\pi^{ij}$ commute with $C_1=p_\lambda$ and $C_2=p_\chi$ the Dirac brackets coincide with the Poisson brackets.


\bibliography{bibliography} 

\end{document}